\documentclass[pdflatex,sn-mathphys-num]{sn-jnl}

\usepackage{tabularx}
\usepackage{placeins}
\usepackage{graphicx}%
\usepackage{multirow}%
\usepackage{amsmath,amssymb,amsfonts}%
\usepackage{bbm}
\usepackage{amsthm}%
\usepackage{mathrsfs}%
\usepackage[title]{appendix}%
\usepackage{xcolor}%
\usepackage{textcomp}%
\usepackage{manyfoot}%
\usepackage{booktabs}%
\usepackage{algorithm}%
\usepackage{algorithmicx}%
\usepackage{algpseudocode}%
\usepackage{listings}%

\theoremstyle{thmstyleone}%
\theoremstyle{thmstyletwo}%

\theoremstyle{thmstylethree}%

\begin{document}

\title[Article Title]{Understanding Interfirm AI Talent Flow Networks through Online Professional Profiles}

\author[1,2,3]{\fnm{Donghang} \sur{Li}}\email{donghang@mit.edu}

\author*[4]{\fnm{Yunhan} \sur{Zheng}}\email{yunhan@pku.edu.cn}

\author[5]{\fnm{Alok} \sur{Prakash}}\email{alok.prakash@smart.mit.edu}

\author[6]{\fnm{Shenhao} \sur{Wang}}\email{shenhaowang@ufl.edu}

\author[2]{\fnm{Jinhua} \sur{Zhao}}\email{jinhua@mit.edu}

\affil[1]{\orgdiv{Department of Civil and Environmental Engineering}, \orgname{Massachusetts Institute of Technology}, \orgaddress{ \city{Cambridge}, \state{MA}, \country{United States}}}

\affil[2]{\orgdiv{Department of Urban Studies and Planning}, \orgname{Massachusetts Institute of Technology}, \orgaddress{ \city{Cambridge}, \state{MA}, \country{United States}}}

\affil[3]{\orgdiv{Department of Electrical Engineering and Computer Science}, \orgname{Massachusetts Institute of Technology}, \orgaddress{ \city{Cambridge}, \state{MA}, \country{United States}}}

\affil*[4]{\orgdiv{College of Urban and Environmental Sciences}, \orgname{Peking University}, \orgaddress{\city{Beijing}, \country{China}}}

\affil[5]{\orgname{Singapore-MIT Alliance for Research and Technology (SMART)}, \orgaddress{ \country{Singapore}}}

\affil[6]{\orgdiv{Department of Urban and Regional Planning}, \orgname{University of Florida}, \orgaddress{\city{Gainesville}, \state{FL}, \country{United States}}}

\abstract{
Artificial intelligence capabilities are often measured as resources accumulated within firms, yet they also circulate across organizational boundaries through worker mobility. Drawing on a database of approximately 535 million employment records across 58 countries, we reconstruct the inter-firm network of AI talent flows over 2010--2022. We find that AI talent inflows are becoming increasingly concentrated among a small set of leading firms, with stronger preferential attachment than in general labor mobility. Yet the network core remains contestable: central positions in AI networks change hands far more often than in non-AI networks. Network position also carries information that goes beyond workforce size.  This network position is economically consequential: AI network centrality is associated with higher enterprise value, after accounting for workforce scale, assets, and fixed effects. Event-study estimates further indicate that sharp improvements in network position are followed by higher enterprise value relative to matched firms with similar pre-event trajectories. Together, these findings suggest that corporate AI advantage depends not only on how much talent firms accumulate, but also on where they sit in its circulation.
}

\keywords{Artificial Intelligence, Labor Flow Network, Human Capital Development}

\maketitle

\section{Introduction}\label{sec1}

Recent advances in artificial intelligence (AI) are rapidly transforming how people learn and work\citep{noy2023experimental,brynjolfsson2025generative,eloundou2024gpts}. As a general-purpose technology with broad applicability across tasks and industries, AI is no longer confined to the technological frontier; it has become a core input into firm strategy, capital allocation and organizational redesign \citep{brynjolfsson2017productivityparadox,agrawal2019economicsai,cockburn2018impactai}. This shift is increasingly visible in the scale of corporate investment and competition. For example, Alphabet projected capital expenditures of \$175--185 billion in 2026, nearly doubling its 2025 spending, as it expanded AI-related infrastructure. \citep{nyt2026alphabet}. At the same time, competition for elite AI talent has intensified: across sectors, the share of job postings requiring AI skills has risen sharply, and AI-related vacancies command substantial wage premia relative to otherwise comparable non-AI positions \citep{alekseeva2021demand}.

Firms are investing aggressively in AI because they expect it to generate future profits, and these expectations are capitalized into market valuations. Although we are still at a very early stage of AI, existing firm-level studies have made substantial progress in quantifying AI-related outcomes. This literature typically models AI as an internal production input proxied by AI workers \citep{babina2024artificial}, AI patents \citep{yang2022artificial}, or AI-related products \citep{nakazato2021artificial}. Although these measures differ, the literature generally conceptualizes AI as an internal firm input: firms hire AI workers, invest in models and infrastructure, and translate these investments into innovation and performance. However, AI capabilities do not accumulate within firms in isolation. They travel with workers across organizational boundaries, giving rise to inter-firm talent networks through which tacit knowledge, frontier practices, and implementation capabilities diffuse.

Recently, network research further suggests that economic outcomes may depend not only on the size of their workforce but also on their structural position in talent mobility networks \citep{nezami2025network}. Central and well-connected actors tend to access information earlier, learn faster, and coordinate resources more efficiently, which can translate into durable performance advantages in competitive environments \citep{li2023mapping}. This inter-firm structure may be especially consequential for AI. AI workers are scarce, expensive, and embedded with rapidly changing technical know-how. When such workers move between firms, they do not merely add headcount; they may also transmit access to valuable knowledge sources \citep{wang2025ai}. This raises a broader question absent from most AI adoption studies: do the economic returns to AI depend only on how many AI workers a firm hires, or also on where those workers come from in the broader talent network?

Therefore, a central gap remains: most studies emphasize the stock of AI talent within firms, while paying limited attention to the networked pathways through which AI capability is transmitted across firms. Labor mobility has been identified as a core transmission mechanism: mobile engineers carry tacit know-how across organizations and regional networks \citep{almeida1999localization}, and firms can ``learn by hiring'' when incoming workers bring experience from knowledge-rich employers \citep{song2003learning}. Taken together, this literature implies that for AI, who firms hire from may matter as much as how many AI workers they hire.

To address these questions, we use Revelio Labs employment records to construct annual inter-firm AI talent mobility networks in the United States, Europe and Southeast Asia over 2010–2022. We define these markets by firm headquarters location and retain only transitions between firms headquartered in the same market. We identify AI roles through a detailed keyword taxonomy, track worker transitions across firms, and construct annual directed weighted networks for AI and non-AI labor flows. We then combine centrality-based network measurement with a panel reduced-form predictive framework that relates future firm performance to current AI employment and AI network position, allowing us to distinguish the predictive content of AI workforce scale from that of network position conditional on scale.

Our results reveal regionally distinct but increasingly concentrated AI talent mobility networks: Big Tech dominates in the United States, AI talent flows are more evenly distributed across industries in Europe, and platforms, finance and telecommunications anchor Southeast Asia. Across all three markets, AI talent flows exhibit stronger preferential attachment than non-AI flows, consistent with cumulative advantage in the AI talent hiring market. Yet rising concentration coexists with continued turnover among core firms, so that central positions remain within reach of firms outside the core. These positions matter for how firms are valued. More central firms have higher subsequent enterprise value than firms with comparable AI workforces, and firms that move sharply up the network gain value relative to matched peers. Together, these findings highlight the concentration, contestability and economic relevance of firms' positions in AI talent mobility networks.

These findings contribute to four literatures. First, this study provides longitudinal comparative evidence on AI worker mobility networks across three major markets, documenting their evolution and regional differences in industrial composition and network structure. Second, we extend firm-level AI adoption research by reframing AI capability as both an internal resource and an inter-firm flow process \citep{brynjolfsson2025generative,eloundou2024gpts,babina2024artificial}. Third, we bring labor-mobility and knowledge-diffusion perspectives into the AI context by characterizing the inter-firm pathways through which AI workers move and the network positions firms occupy \citep{almeida1999localization,song2003learning,rosenkopf2003overcoming,marx2009mobility}. Fourth, we show that firms' positions in AI talent mobility networks provide information about their subsequent market valuation beyond AI workforce scale, with implications for competition policy, talent strategy, and technology governance in the AI era \citep{ahuja2000collaboration,gulati1999alliances,autor2020falling,bloom2020ideas}.

\section{Results}\label{sec2}

\subsection{Constructing and characterizing the AI talent mobility network}

\begin{figure}[!ht]
    \centering
    \includegraphics[width=1\linewidth]{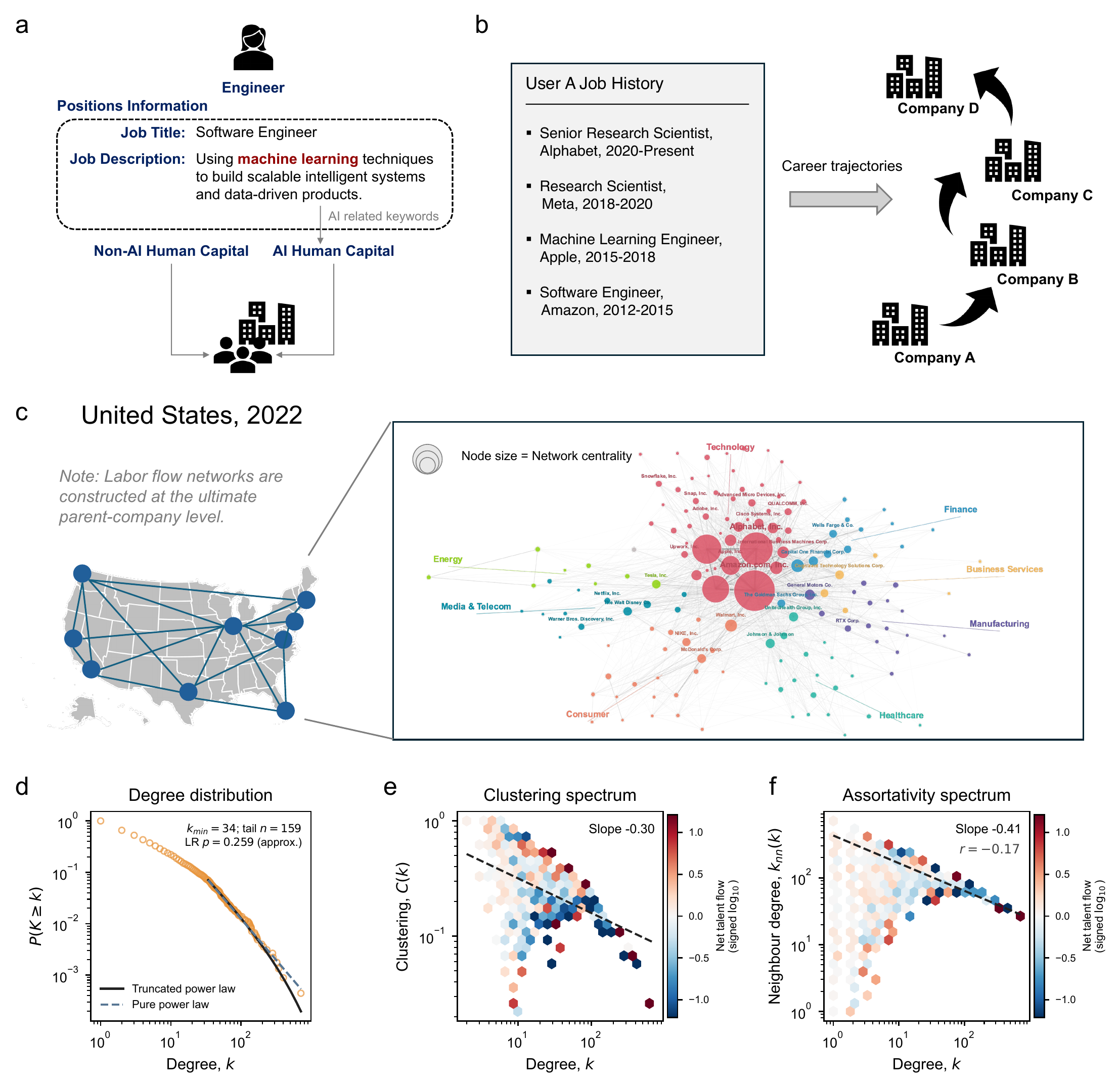}
    \caption{\textbf{Construction and structural characterization of the AI talent mobility network.}
    \textbf{a}, Position-level classification using available job titles, descriptions and AI-related keywords.
    \textbf{b}, Directed transitions between consecutive positions at distinct ultimate-parent firms, dated by the start year of the destination position. AI status is determined by the source position; edge weights count transitions.
    \textbf{c}, Visualization of the United States AI talent mobility network in 2022. Nodes represent ultimate-parent firms, node size encodes PageRank, and colours indicate sectors. The visualization emphasizes leading firms; structural statistics in \textbf{d--f} use all firms observed in the positive-weight edge list, after removing self-loops.
    \textbf{d}, Empirical complementary cumulative degree distribution, $P(K\geq k)$, of the undirected, unweighted projection (2,255 firms). Pure and exponentially truncated power laws are fitted by discrete maximum likelihood on the same tail, $k\geq34$ ($n=159$). Adding an exponential cutoff did not significantly improve the fit (approximate nested likelihood-ratio $p=0.259$).
    \textbf{e}, Local clustering versus degree, with a descriptive log--log fit restricted to positive clustering coefficients (slope $-0.30$).
    \textbf{f}, Mean neighbour degree versus degree, with a degree-weighted log--log fit (slope $-0.41$); degree assortativity is $r=-0.166$. Colours in \textbf{e,f} show mean signed $\log_{10}(1+|\mathrm{net\ flow}|)$ within each hexagonal bin. Further fitting details are provided in the Supplementary Information.}
    \label{fig:ai_mobility_networks}
\end{figure}

To characterize how AI talent moves between firms, we first identified AI-related employment at the level of individual work positions. Using position-level employment records, we classified positions with usable text as AI-related or non-AI-related by matching job titles and descriptions against a curated keyword dictionary (Fig.~\ref{fig:ai_mobility_networks}a; Methods). Positions matching at least one AI-related term were classified as AI-related, whereas those without a match were classified as non-AI-related. These labels describe positions rather than fixed worker categories, allowing a worker's classification to change across their career.

We then reconstructed firm-to-firm mobility from individual career trajectories, assigning each transition according to the AI classification of the source position. Transitions between consecutive positions at different ultimate-parent firms defined directed flows, dated by the start year of the destination position. Aggregating these transitions produced annual directed, weighted networks, with firms as nodes and transition counts as edge weights (Fig.~\ref{fig:ai_mobility_networks}b; Methods). We constructed networks separately for the United States, Europe and Southeast Asia over 2010--2022, retaining flows within each headquarters-defined market. Figure~\ref{fig:ai_mobility_networks}c illustrates the United States AI network in 2022, where large technology and platform firms occupied prominent positions among firms from multiple sectors.

The United States network exhibited three structural signatures in its undirected, unweighted projection, where degree counted distinct partner firms. First, connectivity was highly uneven across the 2,255 firms: the most connected firm had 711 partners, and the degree distribution showed a long right tail (Fig.~\ref{fig:ai_mobility_networks}d). This pattern indicated that a small number of firms connected to many partners, while most firms had relatively few connections. Second, local clustering decreased with degree among firms with positive clustering coefficients (descriptive log--log slope $-0.30$; Fig.~\ref{fig:ai_mobility_networks}e). Thus, the partners of highly connected firms were less densely connected to one another. This pattern was consistent with hubs linking firms across otherwise less interconnected neighbourhoods.
Finally, mean neighbour degree declined with firm degree, alongside negative degree assortativity ($r=-0.166$; Fig.~\ref{fig:ai_mobility_networks}f). Highly connected firms therefore tended to connect to firms with fewer partners, rather than primarily to other hubs. These patterns described an uneven network in which highly connected firms linked many less-connected partners whose connections to one another were comparatively sparse.

This characterization extends the labor-flow network perspective developed for general employment mobility to the circulation of AI-related talent~\citep{guerrero2013employment}. It provides a structural basis for examining how access to AI talent is distributed across firms. We next examine which industries and firms occupy central positions across markets, before assessing how the concentration and persistence of these positions evolve over time.

\subsection{AI talent mobility exhibits distinct regional sectoral patterns}

\begin{figure}[!ht]
    \centering
    \includegraphics[width=1\linewidth]{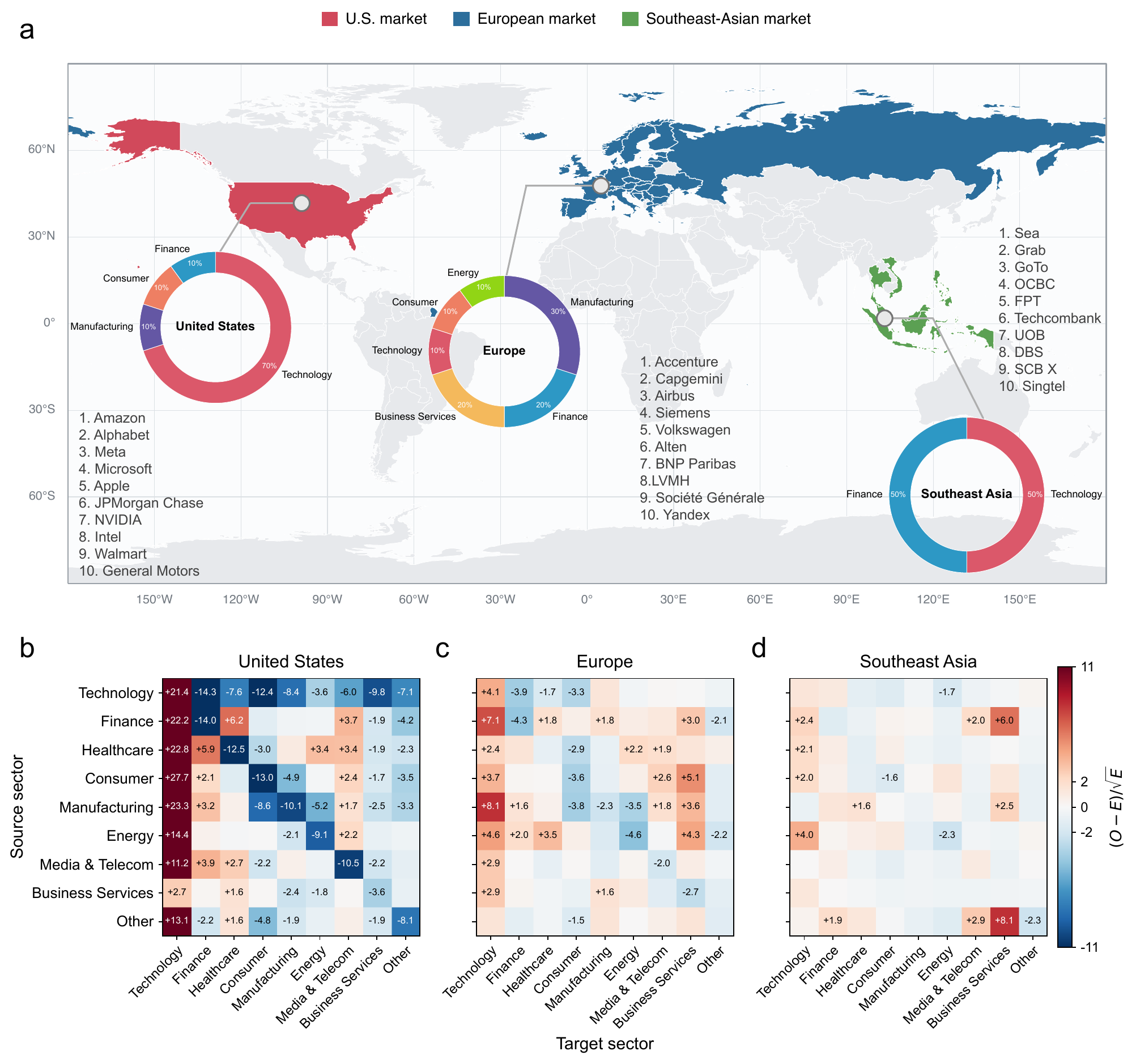}
    \caption{\textbf{Regional sectoral patterns in AI talent mobility networks in 2022.}
    \textbf{a}, Geographic coverage, sector composition, and top PageRank firms in AI talent mobility networks across the United States, Europe, and Southeast Asia. Donut charts summarize the sectoral distribution of the ten most central firms in each market, and numbered lists identify these firms. Sectors follow the classification used in the analysis; Singtel is classified as Technology and Siemens as Energy.
    \textbf{b--d}, Standardized sector-to-sector flow residuals comparing AI and non-AI talent mobility in \textbf{b}, the United States, \textbf{c}, Europe, and \textbf{d}, Southeast Asia. Rows indicate source sectors and columns indicate destination sectors. Expected AI flows preserve each source sector's total AI outflow and allocate destinations according to the corresponding non-AI destination shares (Methods). Positive values indicate flows above this baseline; negative values indicate flows below it. Residuals are descriptive measures of departures from the baseline, rather than tests of statistical significance.}
    \label{fig:industry}
\end{figure}

The industries that anchored AI talent mobility differed across regions, with leading firms spanning distinct combinations of sectors. Ranking firms by PageRank centrality in each market's AI mobility network in 2022 (Fig.~\ref{fig:industry}a) revealed three sectoral profiles. In the United States, technology firms occupied seven of the top ten positions, led by Amazon, Alphabet, Meta, Microsoft, and Apple. JPMorgan Chase, Walmart, and General Motors represented finance, consumer industries, and manufacturing, respectively. In Europe, the leading firms were more broadly distributed: manufacturing accounted for three of the top ten positions, business services and finance for two each, and energy, consumer industries, and technology for one each. Accenture and Capgemini ranked first and second, followed by three manufacturing firms: Airbus, Siemens, and Volkswagen. Southeast Asia showed a different configuration, with the top ten firms divided equally between technology and finance under the sector classification used here. Sea, Grab, and GoTo occupied the first three positions, alongside firms including OCBC, FPT, Techcombank, UOB, DBS, SCBX, and Singtel. Thus, technology dominated the U.S. ranking, whereas the European and Southeast Asian rankings reflected broader industrial and technology--finance configurations, respectively.

To examine how AI-specific mobility departed from non-AI labor mobility, we computed standardized sector-to-sector flow residuals, holding each source sector's total AI outflow fixed and using its non-AI destination shares as the baseline (Fig.~\ref{fig:industry}b--d). The resulting patterns differed across markets. In the United States, flows into technology exceeded the baseline from every source sector. The largest positive residuals linked consumer industries, manufacturing, healthcare, and finance to technology, consistent with cross-sectoral concentration of AI inflows in technology firms. In Europe, the strongest departures were manufacturing-to-technology and finance-to-technology flows, with residuals of 8.1 and 7.1, respectively. Flows from consumer industries, energy, and manufacturing into business services also exceeded the baseline. Although Europe's central firms spanned multiple industries, several flows into technology and business services exceeded the non-AI baseline. Sectoral diversity among central firms therefore coexisted with selective departures from general labor mobility. In Southeast Asia, the largest positive residuals occurred for flows from the heterogeneous Other category and finance into business services (8.1 and 6.0), followed by energy-to-technology flows (4.0). These sector-pair departures complemented the technology--finance composition of its leading firms.

Across the three markets, firm rankings and sectoral flow residuals highlighted complementary dimensions of regional variation. The United States combined technology-dominated centrality with broad excess inflows into technology; Europe combined sectorally diverse leading firms with excess flows into technology and business services; and Southeast Asia combined technology--finance hubs with specific excess flows into business services and technology. These patterns suggest that AI talent mobility is embedded in regional industrial structures, while showing that the sectors containing the most central firms need not coincide with the destinations exhibiting the largest departures from non-AI mobility.

\subsection{AI talent inflows exhibit growing inequality, cumulative advantage, and a dynamic core}

\begin{figure}[!ht]
    \centering
    \includegraphics[width=1\linewidth]{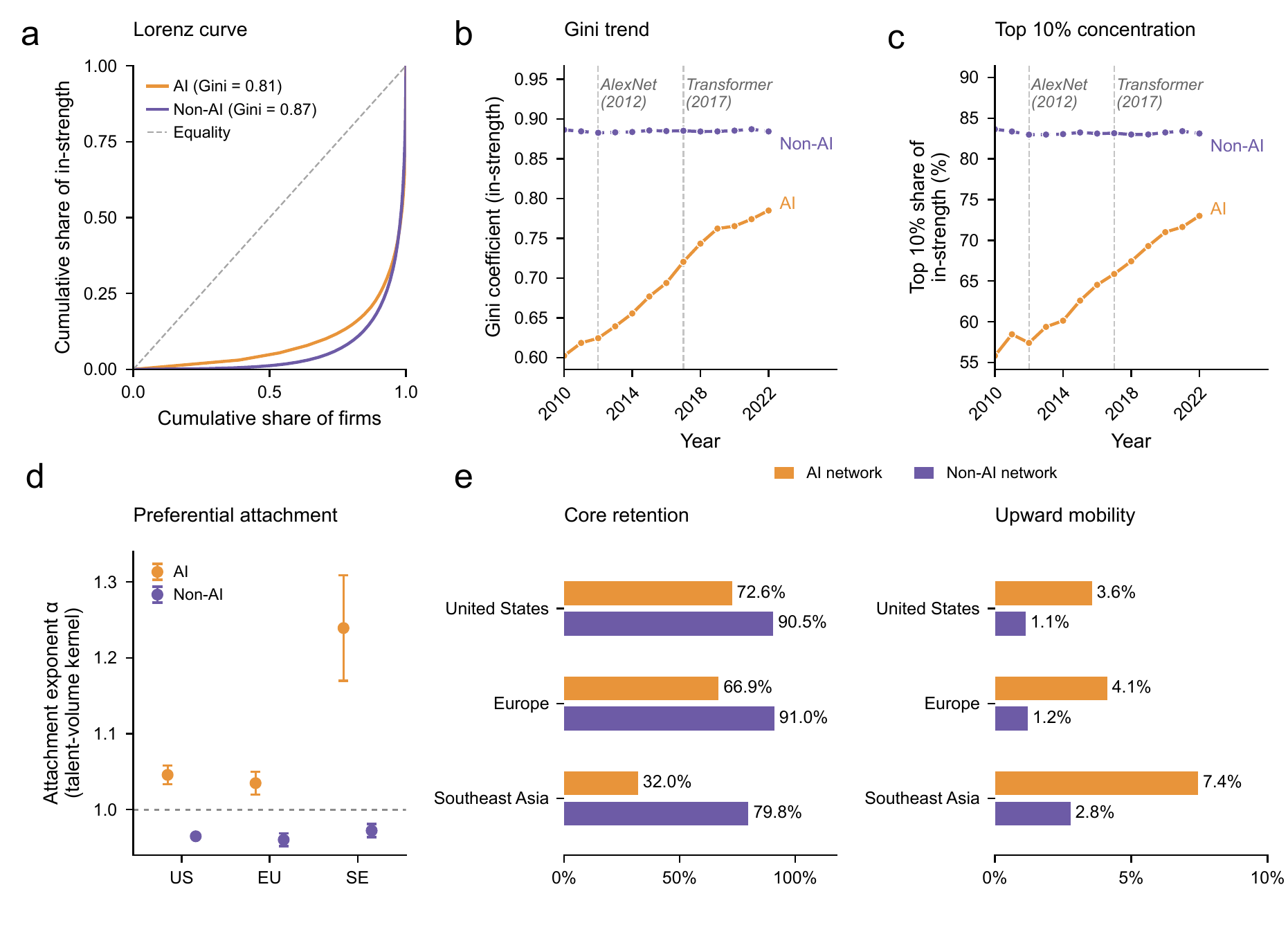}
    \caption{\textbf{Rising concentration of AI talent inflows coexists with turnover in the network core.}
    \textbf{a}, Lorenz curves of firm in-strength in the United States in 2022, for AI and non-AI networks. In-strength measures the volume of incoming talent flows in the given year; curves farther from the equality line indicate greater concentration.
    \textbf{b,c}, Gini coefficients (\textbf{b}) and top-decile shares (\textbf{c}) of annual in-strength, pooling firms across the three markets from 2010 to 2022. Panels \textbf{a--c} include firms with positive in-strength; top-decile shares include ties at the 90th-percentile threshold.
    \textbf{d}, Preferential-attachment exponents for the talent-volume kernel, estimated separately by market and network type using PAFit. Points and vertical bars show estimates and 95\% confidence intervals. The dashed line marks the linear benchmark, $\alpha=1$.
    \textbf{e}, Core retention (left) and upward mobility (right), shown as means of annual transitions from 2010--2011 through 2021--2022. The core comprises firms at or above the 90th percentile of PageRank within each market, network type, and year. Retention is the share of the previous year's core that remains in the core. Upward mobility is the share of the previous year's non-core firms that enters the core; firms first appearing in the current year are excluded from this numerator.}
    \label{fig:network_concentration}
\end{figure}

AI talent inflows became more concentrated across firms over time, narrowing their initial concentration gap with non-AI mobility. Among firms receiving positive inflows, the pooled AI Gini coefficient rose from 0.602 in 2010 to 0.785 in 2022, while the non-AI coefficient remained near 0.885 (Fig.~\ref{fig:network_concentration}b). This narrowing reflected a growing share of AI hiring accruing to leading recipients: firms in the top decile of annual in-strength received 73.0\% of AI inflows in 2022, up from 55.8\% in 2010, whereas the non-AI share remained near 83\% (Fig.~\ref{fig:network_concentration}c). The U.S. Lorenz curves confirmed that a gap persisted in 2022, with Gini coefficients of 0.812 for AI and 0.872 for non-AI inflows (Fig.~\ref{fig:network_concentration}a). AI mobility therefore retained a less concentrated distribution at the end of the period, but an increasing share of its inflows was being absorbed by a relatively small group of firms.

The attachment estimates linked this growing concentration to a pattern of cumulative advantage: larger accumulated AI inflows were associated with more-than-proportional subsequent hiring. PAFit estimates \citep{pham2015pafit} of the talent-volume exponent exceeded one in the United States ($\alpha=1.046$, 95\% CI $[1.033,1.058]$) and Europe ($1.035$, $[1.020,1.050]$), indicating modest departures from proportional attachment in both markets. Southeast Asia showed a larger departure ($1.239$, $[1.170,1.309]$), suggesting that prior hiring advantages were associated with more pronounced differences in subsequent AI inflows there (Fig.~\ref{fig:network_concentration}d). By contrast, non-AI exponents were sub-linear in all three markets ($0.965$, $0.960$, and $0.972$, respectively). Within this model, accumulated hiring volume thus carried a stronger advantage in attracting further AI hires than non-AI hires. This pattern was consistent with cumulative hiring advantage, which may reflect both prior talent inflows and persistent differences in firm attractiveness or hiring capacity.

Concentration of inflows nevertheless coexisted with turnover in the firms occupying central positions. The PageRank-defined AI core retained fewer of its members from one year to the next than the non-AI core: mean retention was 72.6\% versus 90.5\% in the United States, 66.9\% versus 91.0\% in Europe, and 32.0\% versus 79.8\% in Southeast Asia across the 2010--2022 transitions (Fig.~\ref{fig:network_concentration}e, left). Entry from outside the core also remained more common in AI networks, with annual upward-mobility rates of 3.6\%, 4.1\%, and 7.4\%, compared with 1.1\%, 1.2\%, and 2.8\% for non-AI networks (Fig.~\ref{fig:network_concentration}e, right). Hiring advantages therefore did not translate into equally durable core membership. The combination was especially pronounced in Southeast Asia, where the strongest estimated volume attachment coincided with the lowest AI core retention. This distinction between concentrated flows and changing central firms motivated our next question: whether network position provided information about firm performance beyond AI workforce size.

\subsection{Network position carries valuation information beyond AI workforce size}

If a large and growing share of AI talent flows to a small group of firms, but membership of that group keeps changing, a natural question is whether a firm's position in these flows carries information about the firm itself. We asked whether network position helps distinguish how firms are subsequently valued once the size of their AI workforce is taken into account. Our specification follows an AI-augmented production function in which capital, non-AI labor and AI labor enter as inputs and a firm's position in the talent network may shift its efficiency, for example by giving it earlier access to experienced workers and the implementation know-how they carry (Methods, Section~\ref{sec:panel_regressions}). We examined next-year enterprise value and EBITDA. For each outcome, we estimated three models on the same firm-year observations with firm and year fixed effects: one with AI employment only, one with AI PageRank only, and one with AI PageRank together with AI employment, non-AI employment, assets and cash holdings (Table~\ref{tab:baseline_main}). The coefficients therefore describe how changes in a firm's network position relate to changes in its subsequent outcomes. Because PageRank is extremely right-skewed, we entered its logarithm, standardized within each market-year, so that the estimates reflect movements across the whole distribution rather than among a handful of hubs (Methods).

Network position was positively associated with next-year enterprise value, even after accounting for workforce size and other firm characteristics. A one-standard-deviation increase in standardized log AI PageRank was associated with 0.023 log points (about 2.3\%) higher enterprise value in the following year (s.e.\ 0.007; 20,413 firm-years from 3,329 firms; Table~\ref{tab:baseline_main}, column 3). This within-firm association indicates that network position carries information about subsequent valuation beyond AI and non-AI employment, assets and the other included controls. The association with next-year EBITDA was also positive, although smaller and less precisely estimated (0.013 log points; s.e.\ 0.007, $P=0.078$; column 6). One possible interpretation is that AI PageRank reflects a firm's attractiveness and reputation as an employer in the AI talent market \citep{oentaryo2018talent}. By weighting incoming connections according to the network centrality of their source firms, PageRank captures recruitment patterns that workforce counts alone cannot describe. These patterns may signal perceived career opportunities and organizational prospects, offering information relevant to investors’ expectations of future firm performance.

The valuation association was stable across a broad set of alternative specifications (Supplementary Section~8). It changed little when country-by-year or industry-by-year fixed effects absorbed regional and sectoral shocks, when firms whose standardized raw PageRank ever exceeded the pooled 99th-percentile threshold in the baseline regression sample were excluded, when extreme firm sizes were trimmed, when the sample was restricted to firms with AI employment, when leverage and R\&D intensity were added as controls, and when standard errors were clustered by industry or by firm and year. Percentile measures of in-strength and closeness, which also capture embeddedness in the network core, yielded similar associations. The association was present among both technology and non-technology firms and in both the United States and Europe (Supplementary Section~7).

\begin{table}[!htbp]
   \caption{\label{tab:baseline_main}\textbf{AI workforce size, network position and next-year firm performance.}}
   \centering
   \small
   \begin{tabular}{lcccccc}
      \toprule
       & \multicolumn{3}{c}{ln(Enterprise value)} & \multicolumn{3}{c}{ln(EBITDA)}\\
                                & (1)           & (2)           & (3)           & (4)           & (5)           & (6)\\  
      \midrule 
      AI PageRank (log, std.)   &               & 0.059$^{***}$ & 0.023$^{***}$ &               & 0.045$^{***}$ & 0.013$^{*}$\\   
                                &               & (0.010)       & (0.007)       &               & (0.010)       & (0.007)\\   
      Log(1 + AI employees)     & 0.154$^{***}$ &               & -0.052$^{**}$ & 0.175$^{***}$ &               & -0.005\\   
                                & (0.025)       &               & (0.022)       & (0.025)       &               & (0.020)\\   
      Log(1 + non-AI employees) &               &               & 0.258$^{***}$ &               &               & 0.189$^{***}$\\   
                                &               &               & (0.059)       &               &               & (0.056)\\   
      Log assets                &               &               & 0.453$^{***}$ &               &               & 0.587$^{***}$\\   
                                &               &               & (0.034)       &               &               & (0.034)\\   
      Cash / assets             &               &               & -0.126        &               &               & -0.058\\   
                                &               &               & (0.092)       &               &               & (0.106)\\   
      Firms                     & 3,329         & 3,329         & 3,329         & 3,118         & 3,118         & 3,118\\  
       \\
      Observations              & 20,413        & 20,413        & 20,413        & 19,916        & 19,916        & 19,916\\  
      Within R$^2$              & 0.010         & 0.004         & 0.185         & 0.010         & 0.002         & 0.182\\  
       \\
      Firm fixed effects        & Yes           & Yes           & Yes           & Yes           & Yes           & Yes\\  
      Year fixed effects        & Yes           & Yes           & Yes           & Yes           & Yes           & Yes\\  
      \bottomrule
   \end{tabular}
   
   \parbox{\linewidth}{\footnotesize Dependent variables are positive-value logs measured in year $t+1$; regressors are measured in year $t$. AI PageRank is log PageRank standardized within market-year. Columns (1) and (4) include only log(1 + AI employees), columns (2) and (5) only AI PageRank, and columns (3) and (6) add log(1 + non-AI employees), log assets and cash/assets; the three columns for each outcome use the same observations. All models include firm and year fixed effects. Standard errors clustered by firm are in parentheses. $^{*}P<0.10$, $^{**}P<0.05$, $^{***}P<0.01$. }
\end{table}

\FloatBarrier

\subsection{Valuation and firm activity rise after large improvements in network position}

\begin{figure}[!htbp]
    \centering
    \includegraphics[width=1\linewidth]{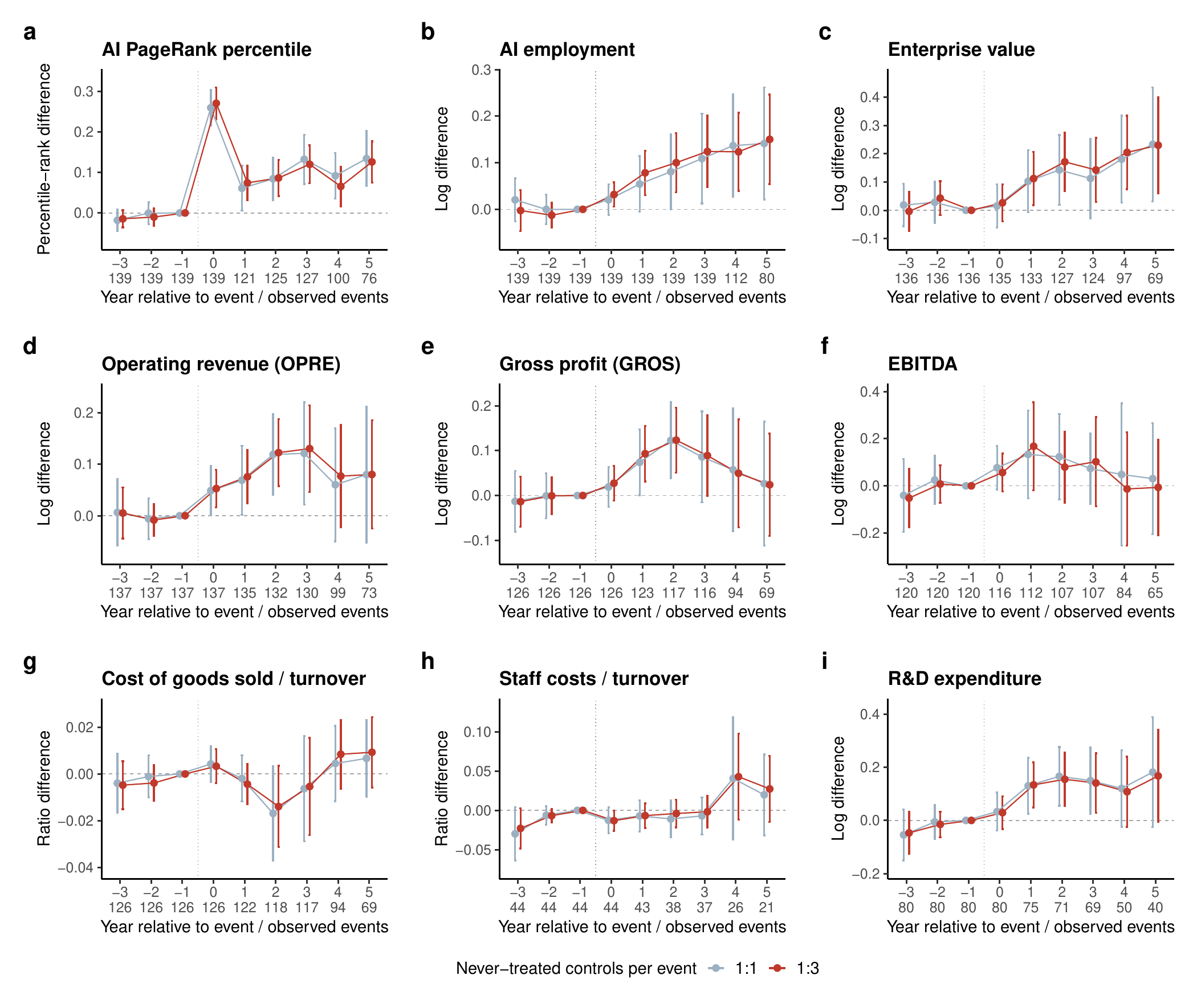}
    \caption{\textbf{Network position, AI employment and firm outcomes around centrality jumps.}
    Stacked event-study coefficients relative to year $-1$ (normalized to zero), with one (light blue) or three (red) matched never-treated controls per event; both use the same event firms and event-time cells. Events are first qualifying jumps in AI PageRank rank in 2013--2019; controls never experienced a qualifying jump and are matched within market, event year and industry on pre-event network position, AI employment, assets and enterprise value (Methods). Models include firm-by-event and event-by-relative-year fixed effects and year-$-1$ log one plus AI employment, log one plus non-AI employment and log assets interacted with relative year. Numbers beneath the year labels give the events observed in each relative year; years 4 and 5 are observed only for events in 2013--2018 and 2013--2017.
    \textbf{a}, Within-market-year AI PageRank percentile rank (139 events).
    \textbf{b}, Log one plus AI employment (139 events).
    \textbf{c}, Log enterprise value (136 events).
    \textbf{d--f}, Log operating revenue (137 events), gross profit (126) and EBITDA (120).
    \textbf{g,h}, Cost of goods sold (126 events) and staff costs (44), each relative to turnover. \textbf{i}, Log R\&D expenditure (80 events).
    Points show estimates and bars show pointwise 95\% confidence intervals clustered by firm. The vertical dotted line separates pre-event and event periods; the horizontal dashed line denotes zero. Joint pre-event $P$ values for panels \textbf{a--i} are 0.235, 0.389, 0.749, 0.907, 0.883, 0.444, 0.800, 0.180 and 0.259 with one control, and 0.397, 0.345, 0.074, 0.758, 0.821, 0.496, 0.572, 0.105 and 0.426 with three controls.}
    \label{fig:event_study_centrality_jump}
\end{figure}

The panel estimates capture associations across annual within-firm changes. To examine outcomes around sharper changes in network position, we compared firms experiencing unusually large increases in PageRank rank with matched firms following similar pre-event trajectories. Events were first qualifying rank increases relative to a firm's maximum rank over the preceding three years. Qualifying increases reached the 90th percentile of such changes during 2013--2019 and were not immediately preceded by a sharp decline. Each event firm was matched to the nearest firm, and in a parallel specification to the three nearest firms, that never experienced a qualifying jump, within the same market, event year and industry. Matching accounted for ranks in each of the preceding three years and pre-event AI employment, assets and enterprise value, and the estimates allow event and control firms of different baseline size to follow different paths. We followed firms for three years before and up to five years after the event (Methods).

The jumps marked sustained improvements in network position. Although the initial gains partly reversed, event firms remained more central than their matched peers throughout the following five years (Fig.~\ref{fig:event_study_centrality_jump}a). AI employment also increased, with larger differences emerging in later years (Fig.~\ref{fig:event_study_centrality_jump}b). Network advances were thus accompanied by an expansion of AI talent capacity.

Enterprise value rose following the jump, with no detectable divergence in valuation trajectories before the event. The valuation difference was small in the event year and generally larger in subsequent years, with a pooled post-event estimate of 0.100 log points (s.e.\ 0.048, $P=0.036$; Fig.~\ref{fig:event_study_centrality_jump}c). Estimates were similar when each event firm was matched to three controls (Supplementary Section~9). This pattern reinforces the panel finding that network position carries information about subsequent valuation: firms experiencing substantial network advances subsequently gained value relative to peers with similar pre-event trajectories. Operating revenue also increased, indicating that the valuation gains were accompanied by an expansion in business activity (Fig.~\ref{fig:event_study_centrality_jump}d). Evidence for higher gross profit and EBITDA was less conclusive (Fig.~\ref{fig:event_study_centrality_jump}e,f), suggesting that this expansion did not yet translate into clearly detectable earnings gains over the observed period.

The expansion in firm activity was not accompanied by a detectable reduction in costs relative to revenue. Neither cost of goods sold nor staff costs relative to turnover declined significantly after the jump (Fig.~\ref{fig:event_study_centrality_jump}g,h), consistent with the previous literature evidence that greater AI investment was not associated with significantly lower cost ratios. Among firms reporting R\&D expenditure, spending increased after the jump, suggesting that network advances may also accompany greater investment in innovation (Fig.~\ref{fig:event_study_centrality_jump}i; Supplementary Section~9). Together, the event-study results complement the panel evidence: substantial improvements in network position were followed by higher valuation and revenue, alongside growth in AI employment and, in the reporting subsample, R\&D spending, but no clear reduction in costs relative to revenue. These changes may reflect a broader expansion of firm capabilities and prospects.

\FloatBarrier

\section{Discussion}\label{sec12}

This study places talent circulation alongside talent accumulation in the study of organizational AI advantage. Research on the economics of AI has emphasized its implications for prediction and innovation \citep{agrawal2019economicsai,cockburn2018impactai}, while firm-level evidence links AI investment to growth and product innovation \citep{babina2024artificial}. Our findings add a relational dimension: firms differ not only in how much AI talent they employ, but also in their positions within its circulation. Across the United States, Europe and Southeast Asia, increasingly concentrated AI talent flows coexisted with turnover among central firms, and network position carried information about subsequent valuation beyond workforce size.

A central finding is that cumulative advantage does not imply a fixed corporate hierarchy. The growing concentration of AI talent inflows connects to evidence that economic activity increasingly accrues to a small group of leading firms \citep{autor2020falling}. Stronger preferential attachment in AI mobility further indicates that firms with larger prior inflows tended to attract disproportionately more subsequent talent. Differences in recruitment resources offer one possible explanation: prior research finds that demand for AI skills is higher among firms with greater market capitalization, cash holdings and R\&D investment \citep{alekseeva2021demand}. Yet the AI network core was less persistent than its non-AI counterpart. Our results therefore distinguish the concentration of talent flows from the persistence of the firms receiving them. A small group can capture an increasing share of talent even as its membership changes, leaving scope for firms outside the core to improve their positions.

These dynamics take different sectoral forms across markets. Big Tech dominates the U.S. network, whereas central firms in Europe span a broader range of industries; Southeast Asia's central firms are concentrated in technology and finance under the classification used here. Labor-flow research shows that worker mobility reflects both geographic and industrial boundaries \citep{park2019global}. Research on engineers likewise demonstrates that the localization of knowledge varies across regional labor networks \citep{almeida1999localization}. The sectoral differences we observe are consistent with AI talent circulation being embedded in these existing structures. Complex economic activities often concentrate geographically \citep{balland2020complex}, and our comparison shows that the industries occupying central positions in AI mobility differ across regional settings.

The valuation results extend research on the economic relevance of firms' network relationships. Studies of embedded exchange emphasize how inter-firm ties shape access to information and coordinated adaptation \citep{uzzi1997social}; alliance research distinguishes resources available through network position from capabilities held within the firm \citep{gulati1999alliances}. Collaboration networks also link direct and indirect connections to innovation outcomes \citep{ahuja2000collaboration}. More recent synthesis finds that the centrality--performance relationship varies with network type and the outcome being measured \citep{nezami2025network}. Our evidence brings this perspective to AI talent mobility: network position was associated with subsequent enterprise value after accounting for workforce size and financial characteristics. The event study reinforces this finding by showing that firms experiencing substantial network advances subsequently gained value relative to peers with similar pre-event trajectories.

One possible source of this association is access to knowledge carried by mobile workers. Learning-by-hiring research shows that recruitment can facilitate knowledge transfer between firms, with its effectiveness depending on the circumstances of the move \citep{song2003learning}. Mobility can also connect firms to knowledge beyond their existing technological and geographic contexts \citep{rosenkopf2003overcoming}. In AI, such connections may provide experience in developing, deploying and integrating technologies that employee counts alone cannot describe. Recent work on AI mobility similarly emphasizes that productive spillovers depend on the organizational context in which workers acquired their expertise \citep{wang2025ai}. Evidence from transfers within multiunit firms further suggests that incoming expertise interacts with the resources of the receiving unit \citep{stadler2022transferring}. These studies motivate a knowledge-access interpretation of our results.

Network position may also convey information about employer attractiveness and organizational prospects. In talent-flow networks, weighted PageRank has been interpreted as indicating the attractiveness of destinations receiving workers from other important nodes \citep{oentaryo2018talent}. Applied to AI mobility, a central position may reflect a firm's ability to attract specialized workers from well-connected employers. Such recruitment patterns could signal career opportunities, reputation or expectations about future growth that are also relevant to investors. This interpretation helps explain why network position may be associated with valuation even when its association with reported earnings is less clear. It also allows for a common underlying source: improving business prospects may attract both talent and investor interest.

The difference between valuation and earnings also connects to the complementary investments required to use AI effectively. The productivity J-curve framework emphasizes that general-purpose technologies require investments in organizational processes, skills and other intangible assets whose benefits may emerge only after an initial period of adjustment \citep{brynjolfsson2021productivity}. Research on the growing burden of knowledge highlights the specialization and teamwork required to develop expertise \citep{jones2009burden}, while evidence of rising research effort underscores the resources needed to sustain innovation \citep{bloom2020ideas}. In this context, access to AI talent may accompany investment in capabilities before yielding clearly measurable earnings gains. Our panel results showed a clearer association with valuation than with EBITDA, and the event study linked network advances to higher revenue and, among reporting firms, greater R\&D spending, without a detectable reduction in cost ratios. Together, these patterns are compatible with capability building and business expansion.

These findings suggest that talent strategy involves both workforce scale and access to the wider labor network. For managers, the origins and pattern of recruitment may help characterize capabilities that headcount alone misses. For policymakers, increasing concentration raises questions about which firms can access specialized talent and the experience it carries. Research on talent discovery shows how uncertainty about ability can restrict opportunities for less-established workers \citep{tervio2009superstars}, while evidence from scientific collaboration demonstrates that expertise can generate benefits beyond the individual possessing it \citep{azoulay2010superstar}. Our results bring attention to the organizational distribution of access to such expertise. Continued turnover within the AI network core also suggests that monitoring concentration alone would miss changes in which firms occupy advantageous positions.

Several limitations define the scope of these interpretations. AI roles are identified from employment text, and online professional profiles may represent firms, sectors and countries unevenly. The analysis covers within-market flows among listed firms in three headquarters-based markets. Firm and year fixed effects and pre-event matching address some sources of heterogeneity, but changes in strategy, business prospects or workforce composition may jointly influence network position and performance. Financial outcomes also do not directly capture knowledge transfer, productivity or innovation outputs. Linking workers' prior experience to subsequent patents, products and technology deployment would help distinguish access to useful knowledge from employer attractiveness and anticipated growth.

The 2010--2022 study period largely precedes the widespread deployment of generative AI, which may change both the demand for specialized workers and how expertise is shared. Evidence from workplace deployment shows that generative AI can improve performance and facilitate learning, particularly among less-experienced workers \citep{brynjolfsson2025generative}. Task-level research also indicates that exposure to large language models extends across a broad range of occupations \citep{eloundou2024gpts}. These developments raise a further question: will AI tools broaden firms' access to expertise, or increase the advantages of employers that already occupy central positions in talent networks? Following both the concentration of flows and turnover among core firms would help reveal how opportunities to build organizational AI capabilities evolve.

\section{Methods}\label{sec11}

\subsection{Data and sample}

Our primary data source is Revelio Labs' database of approximately 535 million employment records across 58 countries, with coverage spanning 2000--2024. Each record contains the worker's job title, employer, start and end dates, and, where available, a job description. Multilingual text was translated into English with the NLLB-200 model (\textit{facebook/nllb-200-distilled-600M}); translation and language-detection procedures are described in Supplementary Section~2. Employers are linked to firms through Revelio's company mapping at the ultimate-parent level, so that all analyses treat subsidiaries of the same corporate parent as one firm. Firm financials (operating revenue, total assets, EBITDA, net income, enterprise value, employment, R\&D expenditure, and cash holdings) come from Bureau van Dijk's Orbis database and are matched to Revelio firms by ISIN and company name, restricting the sample to publicly listed firms. We analyse three markets defined by headquarters location---the United States, Europe, and Southeast Asia---and retain only within-market labor flows. Considering data quality and the period of interest, we select 2010--2022 as the analysis period within the broader database coverage. This window avoids sparse early-period profiles and the deterioration in transition coverage after 2022. The resulting network panel covers 23,596 listed firms, of which 15,391 are matched to Orbis. Regressions use matched firm-years in which the firm appears in the AI network and all model variables are observed (20,413 firm-years from 3,329 firms in the enterprise-value specification). Ratio variables are winsorized at the 1st and 99th percentiles; financial amounts enter as positive-value logarithms.

\subsection{Identifying AI human capital}
\label{sec:identification}

We classify each position as AI-related by matching its job title and description against a curated dictionary of 213 AI-related terms organized in a six-category taxonomy (general technology, predictive AI, generative AI, robotics, automation products, and hardware). The dictionary builds on keyword lists used in prior firm-level AI research \citep{nakazato2021artificial,babina2024artificial}, an approach with documented high precision \citep{miric2023using}; category assignments combine large-language-model reasoning (GPT-4o) with human review. For positions with usable title or description text, matching is case-insensitive and imposes word-boundary constraints around each dictionary term. Positions matching at least one term are classified as AI-related; those with no matches, and those without usable title or description text, are classified as non-AI-related. The six category indicators are assigned independently, allowing a position to receive multiple category labels. The full dictionary, category definitions, and classification rationale are reported in Supplementary Section~3.

\subsection{AI talent mobility networks}

For each year $t$ and worker type $m \in \{\text{AI},\ \text{non-AI},\ \text{all}\}$, we construct a directed, weighted network $G_t^{(m)} = (V_t^{(m)}, E_t^{(m)})$, in which nodes are ultimate-parent firms and a directed edge $i \rightarrow j$ carries weight
\begin{equation}
w_{ij,t}^{(m)} = \text{number of workers of type } m \text{ moving from firm } i \text{ to firm } j \text{ in year } t.
\end{equation}
A transition is recorded when consecutive positions in a worker's employment history belong to different parent firms, dated by the start year of the new position. The AI network uses transitions whose source position, the position the worker leaves, is classified as AI-related; the non-AI network uses all remaining transitions, including those from positions that could not be classified. Each transition therefore enters exactly one of the two networks. Networks are constructed separately by market and year, yielding a dynamic sequence $\{G_t^{(m)}\}_t$ that tracks how firms' network positions evolve across years and worker types.

\subsection{Network position measures}

Our focal measure of network position is weighted PageRank, which scores a firm highly when it receives talent inflows from firms that are themselves central:
\begin{equation}
\text{PR}_{i,t} = \frac{1-\alpha}{|V_t|} + \alpha \sum_{j \in V_t} \text{PR}_{j,t}\, \frac{w_{ji,t}}{s^{out}_{j,t}},
\end{equation}
with damping parameter $\alpha = 0.85$ and out-strength $s^{out}_{j,t} = \sum_{k} w_{jk,t}$. We complement PageRank with in- and out-degree (the number of distinct source and destination firms), in- and out-strength (total inflow and outflow volumes, $s^{in}_{i,t} = \sum_j w_{ji,t}$), net flow ($s^{in}_{i,t} - s^{out}_{i,t}$), and three path- and spectrum-based measures: closeness, betweenness, and eigenvector centrality. Formal definitions of the latter, and the inverse-weight distance transformation they require are given in Supplementary Section~4; all measures are computed on the full annual network, with exact betweenness centrality. In the regressions, PageRank enters as its logarithm standardized to zero mean and unit variance within each market-year, so that coefficients are comparable across markets and years; alternative codings and centrality measures are reported in Supplementary Section~8.

\subsection{Sectoral mixing residuals}

To compare AI mobility with the general structure of labor mobility, we compute Newman-style standardized residuals for sector-to-sector flows. Let $O^{AI}_{ij}$ denote the observed AI flow from source sector $i$ to destination sector $j$, and $P^{NonAI}(j \mid i)$ the destination distribution of non-AI leavers from sector $i$. The expected AI flow preserves each source sector's total AI outflow while assigning destinations according to the non-AI baseline, $E^{AI}_{ij} = \bigl(\sum_{j'} O^{AI}_{ij'}\bigr)\, P^{NonAI}(j \mid i)$, and the residual is
\begin{equation}
D_{ij} = \frac{O^{AI}_{ij} - E^{AI}_{ij}}{\sqrt{E^{AI}_{ij}}}.
\end{equation}
Positive $D_{ij}$ marks sector pairs in which AI mobility exceeds what general labor mobility would predict.

\subsection{Concentration and preferential attachment}

We measure concentration using annual firm in-strength. The Lorenz curves, Gini coefficients, and top-decile shares in Fig.~\ref{fig:network_concentration}a--c use firms with positive in-strength; the time trends pool firms across markets. The top-decile share includes firms at or above the 90th-percentile in-strength threshold, including ties. Separately, we define each market-year network core as firms at or above the 90th percentile of PageRank. Retention is $|C_{t-1}\cap C_t|/|C_{t-1}|$, where $C_t$ is the core in year $t$. Upward mobility is $|P_{t-1}\cap C_t|/|P_{t-1}|$, where $P_{t-1}$ comprises firms present outside the core in the previous year. We report unweighted means over the 12 annual transitions from 2010--2011 through 2021--2022. For preferential attachment, we estimate $A_k\propto k^{\alpha}$ using the regularized PAFit procedure of \citet{pham2015pafit}, separately by market and network type. Each hire is a time-stamped directed edge, and $k$ is the cumulative number of hires a firm has received before the incoming year's events. We estimate attachment without jointly estimating node fitness; $\alpha=1$ is linear and $\alpha>1$ is super-linear attachment. Further details are given in Supplementary Section~4.

\subsection{Panel regressions}
\label{sec:panel_regressions}

The panel specification is motivated by an AI-augmented Cobb--Douglas production function in which capital, non-AI labor and AI labor enter as inputs:
\begin{equation}
Q_{i,t}=A_{i,t}K_{i,t}^{\alpha}L_{i,t}^{\phi}AI_{i,t}^{\theta},
\end{equation}
where $Q_{i,t}$ denotes firm output, $K_{i,t}$ capital, $L_{i,t}$ non-AI labor and $AI_{i,t}$ the stock of AI workers. Network position may enter through $A_{i,t}$ by capturing access to experienced workers, tacit knowledge and implementation know-how transmitted through mobility \citep{song2003learning,wang2025ai}. We express this possibility as
\begin{equation}
\log A_{i,t}=\mu_i+\lambda_t+\beta C_{i,t}+\gamma Z_{i,t}+\varepsilon_{i,t},
\end{equation}
where $C_{i,t}$ is network position, $Z_{i,t}$ collects time-varying firm characteristics, and $\mu_i$ and $\lambda_t$ are firm and year effects. Substitution gives
\begin{equation}
\log Q_{i,t}=\beta C_{i,t}+\theta\log AI_{i,t}+\phi\log L_{i,t}+\alpha\log K_{i,t}+\gamma Z_{i,t}+\mu_i+\lambda_t+\varepsilon_{i,t}.
\end{equation}

This framework guides the control structure of a reduced-form model of subsequent valuation and earnings. We replace contemporaneous output with the outcome in the following year and use log one plus employment to accommodate zero AI employment:
\begin{equation}
\log Y_{i,t+1}=\beta\,C_{i,t}+\theta\log(1+AI_{i,t})+\phi\log(1+L_{i,t})+\alpha\log K_{i,t}+\gamma\,Z_{i,t}+\mu_i+\lambda_t+\varepsilon_{i,t},
\end{equation}
where $Y_{i,t+1}$ is enterprise value or EBITDA in the following calendar year, $C_{i,t}$ is log AI PageRank standardized within market-year, $AI_{i,t}$ and $L_{i,t}$ are AI and non-AI employment, $K_{i,t}$ is total assets, $Z_{i,t}$ is cash over assets, and $\mu_i$ and $\lambda_t$ are firm and year fixed effects. We use log PageRank because raw PageRank is highly right-skewed, so that a standardized raw measure varies mainly among a few hub firms; the logarithm also matches the log-linear form of the production function. For each outcome, the main-text table reports models with AI employment only, with $C_{i,t}$ only, and with all regressors, estimated on the same complete-case sample, from which firms and years observed only once are removed iteratively. Standard errors are clustered by firm, and stars denote two-sided $P<0.10$, $P<0.05$ and $P<0.01$. We report within-$R^2$ as a measure of model fit.

The Supplementary Information reports subgroup estimates by industry, market, firm size and period, with fully interacted models testing whether subgroup coefficients differ. Robustness checks vary the fixed effects (country-by-year and industry-by-year), add a lagged dependent variable, restrict the sample (trimming total assets, requiring at least five panel years, retaining AI-active firms, excluding firms ever above the pooled 99th percentile of standardized raw PageRank in the baseline regression sample), change the end year, recode network position (percentile ranks and alternative centralities), extend the control set and change the level of clustering. Fixed-effect, lagged-outcome, centrality, control-set and clustering comparisons hold the sample fixed across their columns; sample restrictions and end-year checks change the sample by design.

\subsection{Centrality-jump event study}

To examine how firm outcomes evolved around substantial improvements in network position, we used a matched event-study design. Events were firms' first increases in within-market-year AI PageRank rank that reached the 90th percentile of changes during 2013--2019, measured relative to their highest rank in the preceding three years. We excluded rebounds immediately after a sharp decline and followed firms from three years before to up to five years after the event.

Each event firm was matched to the nearest firm that did not experience a qualifying jump during the study period, within the same market, event year and industry. Matching used Mahalanobis distance,
\begin{equation}
d(i,j)=\left[\frac{(Z_i-Z_j)'\Sigma^{-1}(Z_i-Z_j)}{p}\right]^{1/2},
\end{equation}
where $Z$ contains eight standardized pre-event characteristics ($p=8$): PageRank ranks in each of the preceding three years, log one plus AI employment and its two-year change, log assets, and log enterprise value and its two-year change. Employment, assets and valuation levels were measured in the year before the event; $\Sigma$ is the covariance matrix of these characteristics in the eligible reference sample. Matching therefore accounted for both pre-event network trajectories and firm scale and valuation. A parallel specification used up to three nearest controls per event.

We compared event firms with their matched controls using the stacked event-study model
\begin{equation}
y_{i,e,\tau}=\sum_{k\ne-1}\beta_k\,Treat_{i,e}\mathbbm{1}[\tau=k]+\sum_{k\ne-1}\mathbbm{1}[\tau=k]\,X_{i,e}'\gamma_k+\mu_{i\times e}+\lambda_{e\times\tau}+\varepsilon_{i,e,\tau},
\end{equation}
where $e$ indexes matched events, $\tau\in[-3,5]$ is relative year, and $Treat_{i,e}$ identifies the event firm. The coefficients $\beta_k$ measure differences relative to year $-1$, separately for network position, AI employment and financial outcomes. Firm-by-event fixed effects $\mu_{i\times e}$ account for persistent differences between firms, while event-by-relative-year fixed effects $\lambda_{e\times\tau}$ absorb changes common to each matched comparison. The vector $X_{i,e}$ contains pre-event log one plus AI employment, log one plus non-AI employment and log assets, interacted with relative year to allow trajectories to differ with baseline scale. We also estimated a pooled post-event association by replacing the treatment interactions with $\theta Treat_{i,e}\mathbbm{1}[\tau\geq0]$. Standard errors were clustered by firm. This design traces outcomes around large network changes. Matching thresholds, sample requirements, weighting and robustness procedures are detailed in Supplementary Section~9.

\backmatter

\bmhead{Supplementary information}

This article is accompanied by Supplementary Information, which contains the literature review, sample-construction details, the AI keyword dictionary and classification taxonomy, network-centrality definitions and computational details, preferential-attachment estimation, descriptive statistics, full regression tables, and heterogeneity and robustness analyses.

\bmhead{Acknowledgements}

This research is supported by the National Research Foundation (NRF), Prime Minister’s Office, Singapore under its Campus for Research Excellence and Technological Enterprise (CREATE) programme. The Mens, Manus, and Machina (M3S) is an interdisciplinary research group (IRG) of the Singapore MIT Alliance for Research and Technology (SMART) centre.

\bmhead{Data Availability}

The data used in this study were obtained primarily from Revelio Labs and Moody's through institutional subscriptions to the Wharton Research Data Services (WRDS) platform. Access to these proprietary datasets is subject to the licensing terms and conditions of the respective data providers.

\bmhead{Code Availability}

The analyses were conducted using Python and R. The code and scripts used to reproduce the analyses and support the findings of this study are available from the corresponding author upon reasonable request.

\bmhead{Author Contributions Statement}

Donghang Li: Writing--review \& editing, Writing--original draft, Visualization, Methodology, Investigation, Formal analysis, Data curation, Conceptualization; 
Yunhan Zheng: Writing--review \& editing, Supervision, Methodology, Formal analysis, Conceptualization; 
Alok Prakash: Supervision, Project administration; 
Shenhao Wang: Supervision, Project administration; 
Jinhua Zhao: Supervision, Project administration, Funding acquisition.

\bmhead{Competing Interests Statement}

The authors declare no competing interests.

\bibliography{sn-bibliography}

\end{document}


\title[Article Title]{Supplementary Information: Understanding Interfirm AI Talent Flow Networks through Online Professional Profiles}

\author[1,2,3]{\fnm{Donghang} \sur{Li}}\email{donghang@mit.edu}

\author*[4]{\fnm{Yunhan} \sur{Zheng}}\email{yunhan@pku.edu.cn}

\author[5]{\fnm{Alok} \sur{Prakash}}\email{alok.prakash@smart.mit.edu}

\author[6]{\fnm{Shenhao} \sur{Wang}}\email{shenhaowang@ufl.edu}

\author[2]{\fnm{Jinhua} \sur{Zhao}}\email{jinhua@mit.edu}

\affil[1]{\orgdiv{Department of Civil and Environmental Engineering}, \orgname{Massachusetts Institute of Technology}, \orgaddress{ \city{Cambridge}, \state{MA}, \country{United States}}}

\affil[2]{\orgdiv{Department of Urban Studies and Planning}, \orgname{Massachusetts Institute of Technology}, \orgaddress{ \city{Cambridge}, \state{MA}, \country{United States}}}

\affil[3]{\orgdiv{Department of Electrical Engineering and Computer Science}, \orgname{Massachusetts Institute of Technology}, \orgaddress{ \city{Cambridge}, \state{MA}, \country{United States}}}

\affil*[4]{\orgdiv{College of Urban and Environmental Sciences}, \orgname{Peking University}, \orgaddress{\city{Beijing}, \country{China}}}

\affil[5]{\orgname{Singapore-MIT Alliance for Research and Technology (SMART)}, \orgaddress{\country{Singapore}}}

\affil[6]{\orgdiv{Department of Urban and Regional Planning}, \orgname{University of Florida}, \orgaddress{\city{Gainesville}, \state{FL}, \country{United States}}}

\abstract{
Artificial intelligence capabilities are often measured as resources accumulated within firms, yet they also circulate across organizational boundaries through worker mobility. Drawing on a database of approximately 535 million employment records across 58 countries, we reconstruct the inter-firm network of AI talent flows over 2010--2022. We find that AI talent inflows are becoming increasingly concentrated among a small set of leading firms, with stronger preferential attachment than in general labor mobility. Yet the network core remains contestable: central positions in AI networks change hands far more often than in non-AI networks. Network position also carries information that goes beyond workforce size.  This network position is economically consequential: AI network centrality is associated with higher enterprise value, after accounting for workforce scale, assets, and fixed effects. Event-study estimates further indicate that sharp improvements in network position are followed by higher enterprise value relative to matched firms with similar pre-event trajectories. Together, these findings suggest that corporate AI advantage depends not only on how much talent firms accumulate, but also on where they sit in its circulation.
}

\keywords{Artificial Intelligence, Labor Flow Network, Human Capital Development}

\maketitle

\section{Literature review}

Previous studies have documented the heterogeneous effects of artificial intelligence on labor markets and firm performance. Despite this growing body of work, most studies conceptualize AI primarily as a firm's internal stock of human capital or technology, rather than as a capability that circulates across firms through worker mobility. Meanwhile, the literature on network evolution has paid limited attention to systematic differences between AI and non-AI labor flows. This study contributes to four strands of literature: (i) firm-level AI adoption and its economic consequences; (ii) labor mobility as a channel of knowledge diffusion; (iii) inter-firm labor flow networks and their structural properties; and (iv) the concentration of technological capability and its contribution to firm-level and regional inequality.

\subsection{AI adoption and firm performance}

A growing firm-level literature has sought to quantify how artificial intelligence affects productivity, growth, and market valuation. Most of this work treats AI as an internal production input, measured either by the share of AI-skilled workers in a firm's workforce, the stock of AI-related patents, or the deployment of AI-powered products. Researchers document that firms investing more heavily in AI grow faster in sales, employment, and market valuation, and that these gains are concentrated among large and already-leading firms \citep{babina2024artificial}. Complementary evidence from patents \citep{yang2022artificial} and product launches \citep{nakazato2021artificial} reaches broadly similar conclusions: AI adoption is associated with significant firm-level performance gains but is highly unequal across firms. Experimental and task-level studies further show that modern generative AI tools raise worker productivity in knowledge-intensive tasks \citep{noy2023experimental,brynjolfsson2025generative,eloundou2024gpts}, while demand-side evidence reveals a rapidly growing wage premium for AI skills \citep{alekseeva2021demand}. A common feature of this literature is that AI capability is modeled as accumulating within firm boundaries. Our analysis departs from this framing by showing that firms' economic exposure to AI also depends on their position in the inter-firm mobility network through which AI-related human capital, tacit knowledge, and implementation capabilities circulate.

\subsection{Labor mobility and knowledge diffusion}

A second strand of literature emphasizes that labor mobility is a central mechanism through which tacit knowledge travels between organizations. Inventor mobility generates highly localized patterns of knowledge spillovers \citep{almeida1999localization}, and there is direct evidence of ``learning by hiring,'' whereby firms gain access to technological know-how by recruiting engineers from knowledge-rich employers \citep{song2003learning}. Prior studies further show that mobility restrictions materially affect firms' ability to absorb external knowledge and that engineer movements reshape firms' technological trajectories \citep{rosenkopf2003overcoming,marx2009mobility}. This literature implies that for knowledge-intensive technologies such as AI, the origin and structure of hiring relationships may matter in addition to the number of AI workers a firm employs. We operationalize this insight by constructing AI talent mobility networks and examining whether firms' network positions provide information about subsequent performance beyond AI workforce scale.

\subsection{Inter-firm networks and firm performance}

A third strand of literature examines how firms' structural positions in inter-firm networks---whether alliance networks, co-invention networks, supply chains, or labor-flow networks---shape their innovative output and economic performance. Central positions in collaboration networks have been shown to enhance firms' access to information and resources, thereby fostering innovation and growth \citep{ahuja2000collaboration,gulati1999alliances}. More recent work applies similar logic to labor flow networks: workers carry embodied knowledge across firms, so firms repeatedly connected to central employers may enjoy privileged access to frontier capabilities \citep{li2023mapping,nezami2025network}. This perspective is especially relevant for AI, where mobile workers may transmit not only technical skills but also model-building experience, organizational routines, and implementation know-how \citep{wang2025ai}. Centrality measures such as PageRank, in-strength, closeness, and eigenvector centrality capture different forms of embeddedness in the mobility network, while betweenness captures brokerage between otherwise disconnected parts of the network. We apply this toolkit to the AI hiring network and show that a firm's embeddedness in the AI talent mobility network is associated with subsequent performance beyond the scale of its AI workforce.

\subsection{Concentration and technology-driven inequality}

Finally, our findings speak to a broader literature on rising firm concentration and the emergence of superstar firms. A secular decline in the labor share has been linked to the rise of highly concentrated ``superstar'' product markets \citep{autor2020falling}, while the production of new ideas appears to be becoming more difficult and increasingly reliant on scale and complementary investments \citep{bloom2020ideas}. Related work on complex knowledge and regional inequality shows that advanced capabilities tend to cluster disproportionately in leading firms, places, and networks \citep{balland2020complex,park2019global}. Our evidence on the persistent thinness of the AI hiring network, the high concentration of AI in-strength, and the rising top-firm share of AI talent inflows suggests that a similar dynamic is unfolding in the market for AI labor itself. By showing that leading firms are capturing not only larger volumes of AI talent but also more central positions in the networks through which AI-related capabilities circulate, we identify a network-based mechanism through which technological change may translate into firm-level inequality.

\newpage

\section{Data and Sample Construction}
\label{si:sample}

\subsection{Translation and language detection}

Because Revelio position records are multilingual, all job titles and descriptions were mapped to English before classification. Positions from countries whose professional profiles are predominantly English are retained as recorded. For the remaining records, we apply dictionary-based language detection to each job title: titles detected as English are retained, and all other titles (together with their job descriptions) are translated with the NLLB-200 model (\textit{facebook/nllb-200-distilled-600M}), using automatic source-language identification with English as the target language. Translated titles that still fail English detection revert to the original text and are excluded from keyword classification.

\subsection{Markets and final sample}

Firms are assigned to one of three markets by headquarters location: the United States, Europe, and Southeast Asia. Networks are constructed within markets, so cross-market flows are excluded. The underlying employment database spans 2000--2024. We select 2010--2022 as the analysis period considering data quality and the period of interest, excluding sparse early profiles and reduced transition coverage after 2022. Linking to Orbis by ISIN restricts the sample to publicly listed firms. The resulting network panel covers 23,596 listed firms over 2010--2022; the Orbis-matched regression subset varies by outcome (for example, 20,413 firm-year observations from 3,329 firms in the enterprise-value specification).

\section{AI Human Capital Identification}

We identify AI-related positions by matching a manually curated dictionary of AI-related terms against the available job titles and descriptions in the Revelio Labs employment records. Positions without a usable title or description cannot be matched and are treated as non-AI-related, so that every position, and hence every job transition, belongs to exactly one of the two groups. Because AI is an inherently heterogeneous technology, a single keyword list cannot faithfully distinguish between foundational AI research, applied predictive modelling, generative-AI development, robotics, end-user AI products, and the hardware that supports AI computation. We therefore construct a six-category taxonomy that separates core technical fields from task-based and product-based roles. Table~\ref{ai_classification} reports the definitions and illustrative examples for each of the six categories, and Table~\ref{tab:ai-keywords} lists the full dictionary of 213 terms, classified into the corresponding category. Matching is performed at the position level using case-insensitive regular expressions, with word-boundary constraints at the beginning and end of each dictionary term. A position is classified as AI-related if its available job title or description matches at least one term. Category indicators are assigned independently: a position may belong to multiple categories if it matches terms from more than one category. The overall AI-related indicator is binary, so a position matching multiple categories is counted only once in the overall AI measure.

\begin{table}[!htbp]
\centering
\caption{Definitions of AI-related Classification Categories}
\label{ai_classification}
\begin{tabularx}{\textwidth}{lXX}
\hline
\textbf{Category} & \textbf{Definition} & \textbf{Examples} \\
\hline
\textbf{General Technology} & Foundational fields or methods in AI or computing. Refers to broad technical areas, not specific tasks or products. Mutually exclusive with Predictive, Generative, and Robotics. & computer vision, NLP, convolutional neural network (CNN), autoencoder, transformer, artificial intelligence, PyTorch \\
\hline
\textbf{Predictive} & Technologies or models specifically designed to forecast or classify outcomes. & Adaboost, Bayesian network, collaborative filtering, classification algorithms \\
\hline
\textbf{Generative} & Technologies or models that create new content. Task- or model-based only — excludes any end-user products. & adversarial network, BERT, code generation, variational autoencoder (VAE), image generation \\
\hline
\textbf{Robotics} & Technical systems or concepts related to physical robots, including sensing, control, and human-robot interaction. & SLAM, robot navigation, robotic arm control, collaborative system \\
\hline
\textbf{Automation Product} & End-user tools or applications built to automate tasks. This includes AI-powered assistants, chatbots, and coding tools. & ChatGPT, Claude, coding assistant, GitHub Copilot, autonomous vehicle \\
\hline
\textbf{Hardware} & Physical computing components that enable or support AI technologies. & CPU, GPU, computing chips, edge devices \\
\hline
\end{tabularx}
\end{table}

\begin{table*}[!htbp]
\centering
\small
\caption{Keyword list used to identify AI-related employment positions, classified into six functional categories. Available job titles and descriptions are matched using case-insensitive regular expressions with word boundaries at the beginning and end of each dictionary term.}
\label{tab:ai-keywords}
\renewcommand{\arraystretch}{1.25}
\begin{tabularx}{\textwidth}{@{}p{2.7cm}X@{}}
\toprule
\textbf{Category} & \textbf{Keywords} \\
\midrule

General Technology \mbox{(95 terms)} &
AGI, AI, ANI, artificial general intelligence, artificial intelligence, artificialintelligence,
augmented reality, autoencoder, autonomic computing, Backpropagation, cognitive computing,
computational intelligence, computational linguistics, computer vision, connectionism,
connectionist, convolutional neural network, core software system, creative coding,
deep belief network, deep learning, DL, evolutionary computation, expert system, fuzzy c,
fuzzy environment, fuzzy logic, fuzzy number, fuzzy set, fuzzy system, high-dimensional data,
high-dimensional feature, high-dimensional input, high-dimensional model, high-dimensional space,
high-dimensional system, high-performance compute, inductive logic programming, intelligent agents,
Keras, kernel methods, latent variable, learning algorithm, learning model, Libsvm,
long short term memory, Long Short Term Memory, Long Short-Term Memory, LSTM, machine intelligence,
machine learning, machine translation, MapReduce, ML, multi-agent system, multi-layer perceptron,
multi-objective optimisation, MXNet, natural gradient, natural language processing, neural network,
neural networks, neural turing, neuromorphic computing, NLP, non negative matrix factorisation,
object detection, object recognition, OpenCV, OpenNLP, pattern recognition, policy gradient methods,
pytorch, Q-learning, Recurrent Neural Network, reinforcement learning, RL, RLHF, RNN, scikit-learn,
semi-supervised learning, sparse coding, sparse representation, speech recognition,
stochastic gradient, supervised learning, swarm behavior, swarm intelligence, TensorFlow, Theano,
transfer learning, unsupervised learning, variational inference, virtual reality,
voice recognition, word2vec. \\
\addlinespace[3pt]
\midrule

Predictive AI \mbox{(58 terms)} &
Adaboost, ant colony, association rule, bayesian network, bayesian networks, bee colony,
classification algorithms, collaborative filtering, DBSCAN, decision model, decision tree,
differential evolution algorithm, dimensionality reduction, ensemble learning, evolutionary algorithm,
factorisation machine, feature engineering, feature extraction, feature learning, feature selection,
firefly algorithm, fraud detection, gaussian mixture model, gaussian process, genetic algorithm,
genetic programming, gradient boosting, gradient tree boosting, hidden Markov model, image recognition,
Information Extraction, K-means, kernel learning, latent dirichlet allocation, latent semantic analysis,
link prediction, logistic regression, memetic algorithm, multi-label classification,
multi-objective evolutionary algorithm, naive bayes, particle swarm optimisation, predictive analytics,
predictive modeling, random forest, Random Forests, rankboost, recommender system, Recommender Systems,
Sentiment Analysis, spectral clustering, support vector machine, Support Vector Machines, SVM,
topic model, vector machine, xgboost. \\

\bottomrule
\end{tabularx}
\end{table*}

\begin{table*}[!htbp]
\centering
\small
\caption[]{Keyword list used to identify AI-related employment positions, classified into six functional categories (\textit{continued}).}
\renewcommand{\arraystretch}{1.25}
\begin{tabularx}{\textwidth}{@{}p{2.7cm}X@{}}
\toprule
\textbf{Category} & \textbf{Keywords} \\
\midrule

Generative AI \mbox{(25 terms)} &
adversarial network, audio generation, BERT, code generation, content generation,
creative adversarial network, creative ai, creative artificial intelligence, DALLE, diffusion model,
diffusion models, GAN, GenAI, generative adversarial network, generative ai,
generative artificial intelligence, generative model, GPT, image generation, natural language generation,
stable diffusion, style transfer, text generation, VAE, variational autoencoder. \\
\addlinespace[3pt]
\midrule

Robotics \mbox{(7 terms)} &
collaborative system, collaborative systems, robot, robotic, robotics, robots, SLAM. \\
\addlinespace[3pt]
\midrule

Automation Products \mbox{(20 terms)} &
automated vehicles, autonomous driving, autonomous vehicle, autonomous vehicles, bard, chat gpt,
chatbot, chatbots, ChatGPT, claude, coding assistant, copilot, deepseek, gemini, midjourney,
notion ai, notionai, self-driving cars, sora, virtual assistants. \\
\addlinespace[3pt]
\midrule

Hardware \mbox{(8 terms)} &
central processing unit, computing chips, FPGA, GPU, graphics processing unit, integrated circuit,
processor chip, semiconductor. \\

\bottomrule
\end{tabularx}

\vspace{2pt}
\footnotesize\textit{Note:} The full keyword list comprises 213 terms. Each term is mapped to exactly one
category. Matching against available job titles and descriptions uses case-insensitive regular expressions
with word boundaries at the beginning and end of each dictionary term; capitalisation in this table
follows the original entry in the dictionary file.
\end{table*}

\newpage

\section{Labor Flow Network}

\subsection{Structural characterization of the 2022 United States AI network}
\label{si:network_structure}

The structural panels in main Fig.~1 use the 2022 United States AI transition network. We retain positive-weight edges between distinct ultimate-parent firms and construct an undirected, unweighted projection. Reciprocal flows define a single connection, and degree counts distinct partner firms. The graph contains 2,255 edge-observed firms; firms absent from the edge list are not added as zero-degree nodes. The maximum observed degree is 711. This projection is used for the structural diagnostics, whereas the mobility networks retain direction and weights for the other analyses.

We compare a discrete pure power law, $p(k)\propto k^{-\alpha}$, with an exponentially truncated alternative, $p(k)\propto k^{-\alpha}\exp(-\lambda k)$, on infinite integer support above a common threshold. Minimizing the Kolmogorov--Smirnov distance for the pure-power-law fit selects $k_{\min}=34$, retaining 159 firms (7.05\% of the graph). Discrete maximum-likelihood estimation gives $\alpha=2.59695$ for the pure model and $\alpha=2.37794$, $\lambda=0.00150138$ for the truncated model. Fitted conditional survival curves are multiplied by $159/2255$ for comparison with the full empirical distribution and are not extrapolated below the threshold.

The likelihood-ratio statistic is 1.27517, with an approximate nested $\chi^2_1$ comparison giving $p=0.25880$. This comparison is conditional on the selected threshold and does not adjust for threshold selection, dependence among degrees within one graph, or the boundary at $\lambda=0$. Adding an exponential cutoff did not significantly improve the fit.

Local clustering is calculated without edge weights and is fitted against degree only for firms with positive clustering coefficients. Ordinary least squares in log--log coordinates gives a slope of $-0.30129$. The mean-neighbour-degree spectrum is fitted in log--log coordinates using degree as the residual weight in \texttt{numpy.polyfit}, giving a slope of $-0.41415$; its least-squares objective therefore weights squared residuals by degree squared. These slopes summarize the observed relationships. Degree assortativity of the same graph is $-0.16607$. These diagnostics characterize heterogeneous, disassortative connectivity.

\subsection{Network Dynamics}

\begin{figure}[!htbp]
    \centering
    \includegraphics[width=0.95\linewidth]{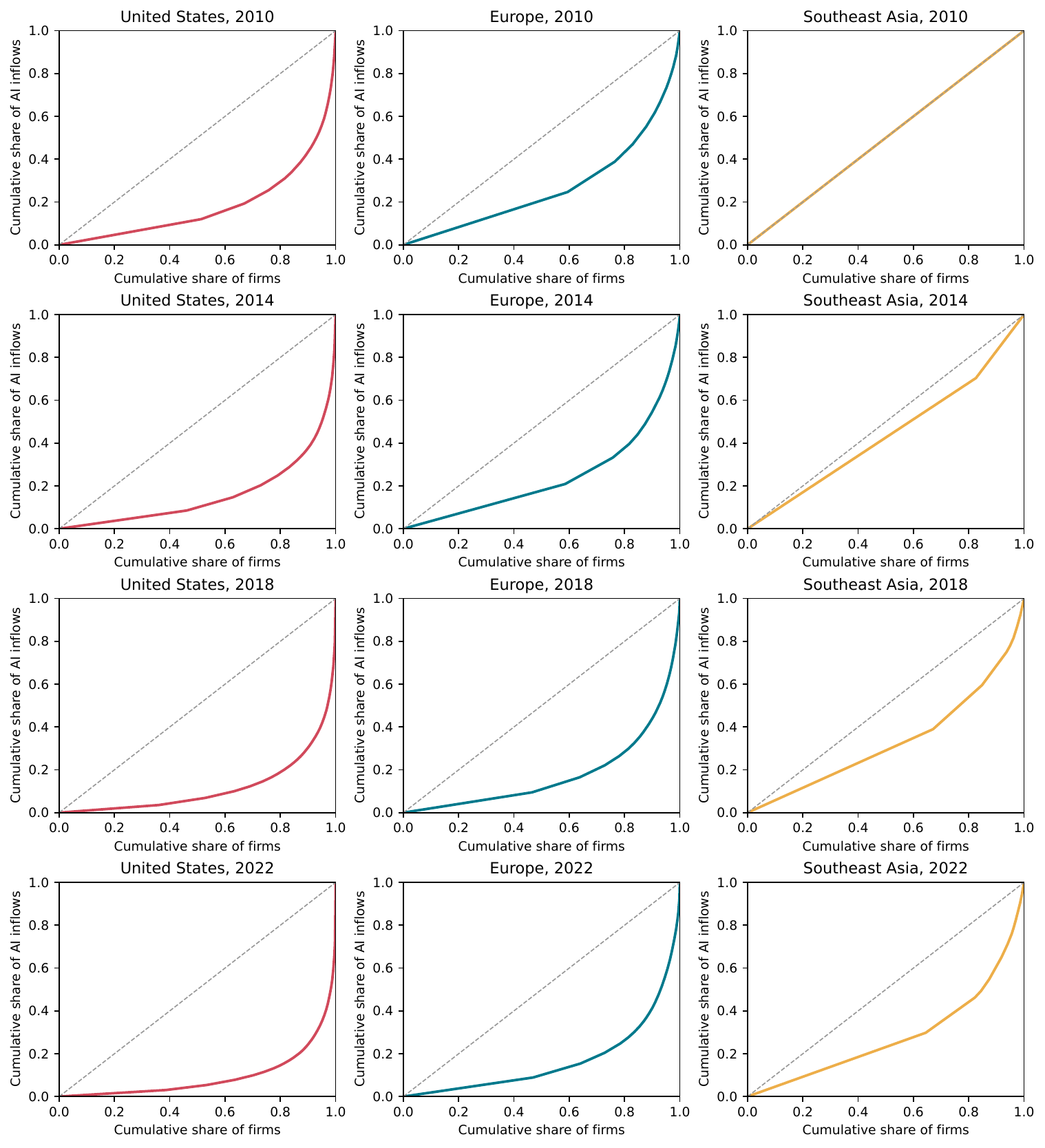}
    \caption{\textbf{Lorenz curves of firm-level AI in-strength across markets and years.} Each panel plots the cumulative share of AI hiring inflows (vertical axis) received by the bottom $x\%$ of firms ranked by AI in-strength (horizontal axis), for the United States, Europe, and Southeast Asia in 2010, 2014, 2018, and 2022. The sample includes firms with positive AI inflows. The 45-degree dashed line corresponds to perfect equality; the Southeast Asia curve in 2010 coincides with this line. Across the four displayed years, the curves move further below the diagonal in each market, indicating increasing concentration among recipient firms.}
    \label{fig:lorenz_panel}
\end{figure}

Figure~\ref{fig:lorenz_panel} presents Lorenz curves of firm-level AI in-strength for the United States, Europe, and Southeast Asia in 2010, 2014, 2018, and 2022, including only firms with positive AI inflows. Each curve plots the cumulative share of AI inflows received by firms ranked from lowest to highest in-strength. The curves lie progressively further below the equality line across these four years in each market, showing increasing concentration among firms receiving AI talent. Southeast Asia begins at equality in 2010, when all eight firms with positive inflows receive the same amount. By 2022, the bottom 80\% of firms receive approximately 14.6\% of AI inflows in the United States, 26.2\% in Europe and 44.2\% in Southeast Asia. These curves complement the pooled main-text concentration statistics by showing how inflow concentration evolves within each market.

\subsection{Computation of network measures}
All network measures are computed on the full annual network for each market and network type, without restricting the graph to a core of heavily weighted edges. Edges with non-positive weights and self-loops are removed, and repeated firm pairs are aggregated. Path-based measures (closeness and betweenness) use the inverse-weight distance $d_{ij,t}=1/w_{ij,t}$, so that larger flows correspond to shorter distances, and betweenness centrality is computed exactly from all source firms.

\subsection{Network Centrality Measures}
\label{si:centrality}

This section provides formal definitions for the centrality measures summarized in the main-text Methods. Let $G_t$ be the directed weighted graph in year $t$ with weight matrix $\mathbf{W}_t = [w_{ij,t}]$.

\paragraph{Degree and strength.}
The in-degree $k^{in}_{i,t} = |\{ j : w_{ji,t} > 0 \}|$ counts distinct source firms, and the out-degree $k^{out}_{i,t} = |\{ j : w_{ij,t} > 0 \}|$ counts distinct destination firms. In-strength and out-strength aggregate flow volumes, $s^{in}_{i,t} = \sum_j w_{ji,t}$ and $s^{out}_{i,t} = \sum_j w_{ij,t}$, and net flow is their difference.

\paragraph{Distance transformation.}
Path-based measures require distances rather than intensities. We transform flow weights into edge distances
\begin{equation}
d_{ij,t} = \frac{1}{w_{ij,t}},
\end{equation}
so that larger hiring flows correspond to shorter effective distances and shortest paths preferentially traverse strong flows.

\paragraph{Closeness centrality.}
Let $R^{in}_{i,t}$ contain the other firms with a directed path to firm $i$, and let $r_{i,t}=|R^{in}_{i,t}|$. We use inward closeness with the Wasserman--Faust correction for the share of firms that can reach $i$:
\begin{equation}
C^{close}_{i,t} = \frac{r_{i,t}}{|V_t|-1}\,\frac{r_{i,t}}{\sum_{j \in R^{in}_{i,t}} \text{dist}_t(j,i)},
\end{equation}
where $\text{dist}_t(j,i)$ is the shortest-path distance from $j$ to $i$, computed using edge distances $d_{ij,t}=1/w_{ij,t}$. The score is zero when no other firm can reach $i$, or when the graph contains only one firm. Higher values indicate shorter incoming paths and a larger share of firms able to reach the destination.

\paragraph{Betweenness centrality.}
With $\sigma_{sr,t}$ the number of shortest paths from firm $s$ to firm $r$ and $\sigma_{sr,t}(i)$ the number passing through firm $i$,
\begin{equation}
C^{betw}_{i,t} = \frac{1}{(|V_t|-1)(|V_t|-2)}\sum_{\substack{s \neq i,\ r \neq i,\ s \neq r}} \frac{\sigma_{sr,t}(i)}{\sigma_{sr,t}}.
\end{equation}
The sum runs over ordered pairs with a directed path, using inverse-weight distances. Scores are normalized by the number of possible ordered pairs excluding $i$, and are zero for graphs with fewer than three firms. High-betweenness firms act as bridges between otherwise weakly connected parts of the labor-flow network.

\paragraph{Eigenvector centrality.}
Eigenvector centrality solves $\mathbf{x}_t = \lambda_t^{-1} \mathbf{A}_t^{\top} \mathbf{x}_t$, where $\mathbf{A}_t$ has entries $a_{ij,t}=w_{ij,t}$ and $\lambda_t$ is its dominant eigenvalue. Scores are normalized to unit Euclidean norm. This incoming-flow convention assigns high scores to firms receiving talent from other high-scoring firms.

\paragraph{Implementation.}
PageRank uses damping parameter $\alpha = 0.85$. All measures are computed on the full annual network, and betweenness centrality is computed exactly.

\begin{table}[!htbp]
\centering
\caption{Network Metrics in the Hiring Flow Network}
\label{tab:network-metrics}
\small
\begin{tabularx}{\textwidth}{l l X}
\toprule
\textbf{Metric} & \textbf{Notation} & \textbf{Definition and interpretation} \\
\midrule

In-degree & $k^{in}_{i,t}$ &
Number of distinct source firms that send at least one worker to firm $i$ in year $t$:
$k^{in}_{i,t} = \left|\{j: w_{ji,t} > 0\}\right|$. Captures diversity of hiring sources. \\

Out-degree & $k^{out}_{i,t}$ &
Number of distinct destination firms that receive at least one worker from firm $i$ in year $t$:
$k^{out}_{i,t} = \left|\{j: w_{ij,t} > 0\}\right|$. Captures breadth of talent outflows. \\

In-strength & $s^{in}_{i,t}$ &
Total number of hires into firm $i$ in year $t$:
$s^{in}_{i,t} = \sum_{j} w_{ji,t}$.
Measures hiring intensity (inflow volume). \\

Out-strength & $s^{out}_{i,t}$ &
Total number of worker outflows from firm $i$ in year $t$:
$s^{out}_{i,t} = \sum_{j} w_{ij,t}$.
Measures the scale of talent outflows. \\

Net flow & $\text{NetFlow}_{i,t}$ &
Difference between inflow and outflow volumes:
$\text{NetFlow}_{i,t} = s^{in}_{i,t} - s^{out}_{i,t}$.
Positive values indicate net talent inflow. \\

PageRank & $\text{PR}_{i,t}$ &
Weighted PageRank computed on the directed network using $w_{ij,t}$ as edge weights.
Captures a firm's importance by accounting for both the magnitude and origin of inflows
(i.e., inflows from central firms contribute more). \\

Closeness centrality & $C^{close}_{i,t}$ &
Inverse of the average incoming shortest-path distance, adjusted for the share of firms able to reach $i$:
$C^{close}_{i,t}=\frac{r_{i,t}}{|V_t|-1}\frac{r_{i,t}}{\sum_{j\in R^{in}_{i,t}}\text{dist}_t(j,i)}$.
Here $R^{in}_{i,t}$ and $r_{i,t}$ are defined above; edge distances are $d_{ij,t}=1/w_{ij,t}$. \\

Betweenness centrality & $C^{betw}_{i,t}$ &
Normalized sum of shortest-path shares through firm $i$ over ordered pairs of other firms:
$C^{betw}_{i,t}=\frac{\sum_{s\neq i,\,r\neq i,\,s\neq r}\sigma_{sr,t}(i)/\sigma_{sr,t}}{(|V_t|-1)(|V_t|-2)}$.
Only reachable pairs contribute. Shortest paths use $d_{ij,t}=1/w_{ij,t}$; computed exactly over all source firms. \\

Eigenvector centrality & $C^{eig}_{i,t}$ &
A recursive measure assigning higher scores to firms connected to other central firms:
$\mathbf{x}_t = \lambda_t^{-1}\mathbf{A}_t^{\top}\mathbf{x}_t$, where $A_{ij,t}=w_{ij,t}$ and $\|\mathbf{x}_t\|_2=1$.
Scores reflect incoming links from other high-scoring firms. \\

\bottomrule
\end{tabularx}
\begin{tablenotes}[flushleft]
\footnotesize
\item \textit{Notes:} $w_{ij,t}$ denotes the number of workers moving from firm $i$ to firm $j$ in year $t$.
All measures are computed on the full positive-weight network for each market, network type and year.
\end{tablenotes}
\end{table}

\subsection{Cross-Market Comparison}

\begin{figure}[!ht]
    \centering
    \includegraphics[width=0.8\linewidth]{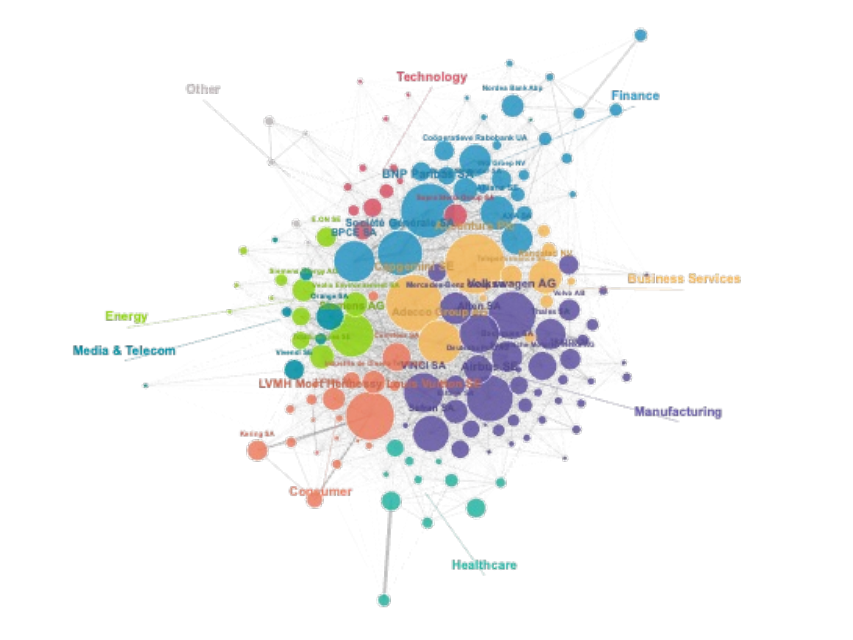}
    \caption{\textbf{Inter-firm AI talent mobility network in Europe, 2022.} Nodes represent firms, with node size proportional to network centrality and node color indicating industry. Directed edges represent observed movements of AI professionals between firms.}
    \label{fig:network_eu}
\end{figure}

\begin{figure}[!ht]
    \centering
    \includegraphics[width=0.8\linewidth]{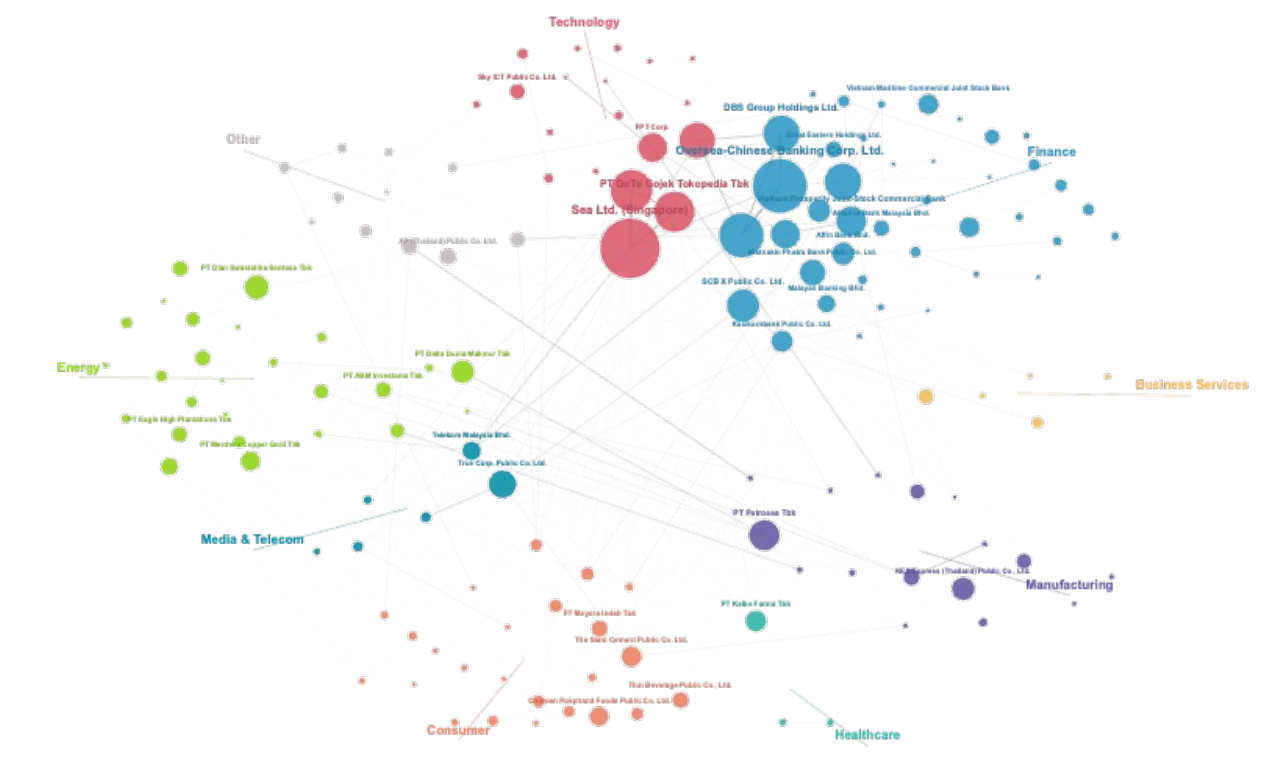}
    \caption{\textbf{Inter-firm AI talent mobility network in Southeast Asia, 2022.} Nodes represent firms, with node size proportional to network centrality and node color indicating industry. Directed edges represent observed movements of AI professionals between firms.}
    \label{fig:network_se}
\end{figure}

Figures~\ref{fig:network_eu} and~\ref{fig:network_se} show distinct regional configurations of AI talent mobility. In the 2022 European AI network, the ten highest-PageRank firms comprised three manufacturing firms, two business-services firms, two financial firms, and one firm each in energy, consumer industries, and technology. In Southeast Asia, five of the ten highest-PageRank firms were classified as technology and five as finance. These categories follow the sector mapping used in the analysis: Siemens was classified as Energy, and Singtel as Technology. Firm rankings describe the sectoral composition of leading firms. Sector-to-sector residuals in main Fig.~2 compare AI flows with source-sector-specific non-AI destination shares.

\begin{figure}[!ht]
    \centering
    \includegraphics[width=1\linewidth]{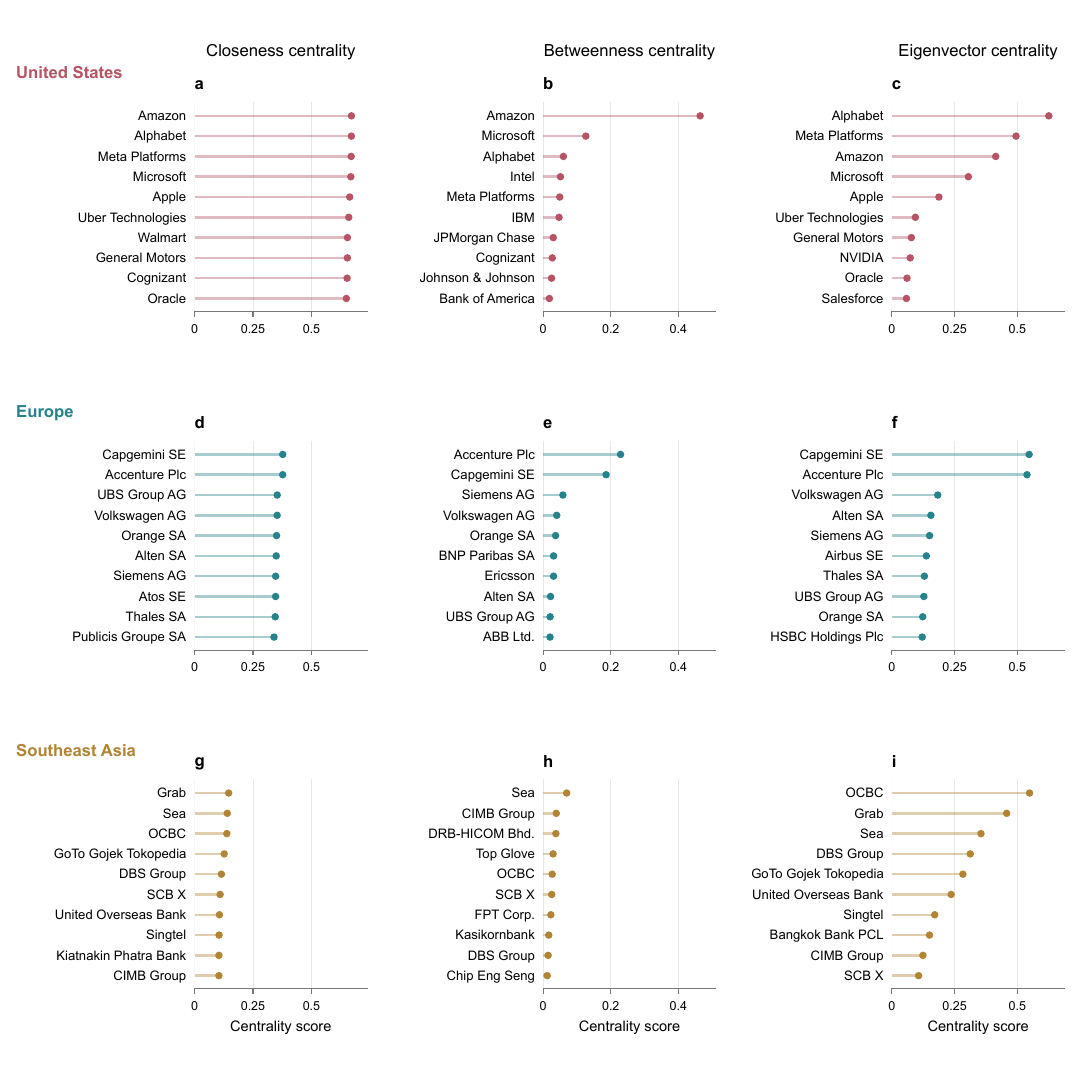}
    \caption{\textbf{Top firms by AI network centrality across markets.} Panels report the leading firms in the United States, Europe, and Southeast Asia according to closeness, betweenness, and eigenvector centrality.}
    \label{fig:centrality_rankings}
\end{figure}

Figure~\ref{fig:centrality_rankings} further demonstrates that firm positions in AI talent mobility networks are multidimensional. Closeness and eigenvector centrality identify firms that are deeply embedded in the network core and closely connected to other influential organizations, whereas betweenness centrality captures firms that broker talent flows between otherwise weakly connected sectors or clusters. Firms that rank highly under one measure do not necessarily rank highly under another, indicating that core embeddedness and brokerage represent distinct forms of network advantage.

These network roles also vary systematically across regions. In the United States, closeness and eigenvector rankings are dominated by major technology firms, including Amazon, Alphabet, Microsoft, Meta, and Apple, indicating that the US AI labor market is centered on a concentrated group of Big Tech firms that function as both major attractors and sources of AI talent. Betweenness rankings are more heterogeneous, however, and include large firms from finance, retail, manufacturing, and business services, suggesting that cross-sector brokerage extends beyond frontier technology companies.

Europe exhibits a broader sectoral composition. Financial institutions such as BNP Paribas, BPCE, AXA, Allianz, and HSBC occupy prominent positions, alongside consulting and IT service firms such as Accenture and Capgemini and advanced manufacturers including Airbus, Mercedes-Benz, Volkswagen, and Daimler Truck. This pattern reflects a more distributed ecosystem in which AI capabilities circulate across finance, professional services, and manufacturing rather than being concentrated primarily within technology firms.

Southeast Asia displays a third configuration, led by digital platforms, regional banks, and telecommunications providers. Platform companies such as Sea, Grab, and GoTo rank prominently across multiple centrality measures, together with financial institutions including DBS, UOB, OCBC, and CIMB and telecommunications firms such as Singapore Telecommunications. The regional network therefore appears to be structured around platform ecosystems, financial intermediation, and digital infrastructure. Taken together, these comparisons show that firms may occupy central positions through different mechanisms and that the economic value of AI talent mobility may depend on whether a firm is embedded in the network core or serves as a broker across sectors and submarkets.

\newpage

\subsection{Preferential Attachment Estimation}
\label{si:pafit}

To quantify cumulative advantage in talent inflows, we estimate the attachment kernel $A_k\propto k^{\alpha}$ separately for each market and network type (AI, non-AI) over 2010--2022. For the talent-volume kernel, each hire is represented as a time-stamped directed edge, so the receiving firm's degree in the growing multigraph counts cumulative incoming hires. The attachment events in year $t$ are related to cumulative hires received from 2010 through $t-1$. We use the regularized PAFit estimator of \citet{pham2015pafit}, implemented with \texttt{only\_A\_estimate} and \texttt{MLE = FALSE}; node fitness is not jointly estimated. The exponent $\alpha=1$ denotes linear attachment, $\alpha>1$ super-linear attachment, and $\alpha<1$ sub-linear attachment. Confidence intervals are those returned by PAFit.

The talent-volume estimates in main Fig.~3d are $1.0458$ (95\% CI $[1.0334,1.0582]$), $1.0350$ ($[1.0200,1.0499]$), and $1.2393$ ($[1.1696,1.3089]$) for AI networks in the United States, Europe, and Southeast Asia, respectively. The corresponding non-AI estimates are $0.9649$ ($[0.9623,0.9675]$), $0.9604$ ($[0.9520,0.9687]$), and $0.9724$ ($[0.9637,0.9811]$). Thus, the AI confidence intervals lie above one in all three markets, whereas the non-AI intervals lie below one.

\section{Distribution of AI Talent Inflows}

\subsection{Descriptive CCDF analysis}

\begin{figure}[!htbp]
    \centering
    \includegraphics[width=0.95\linewidth]{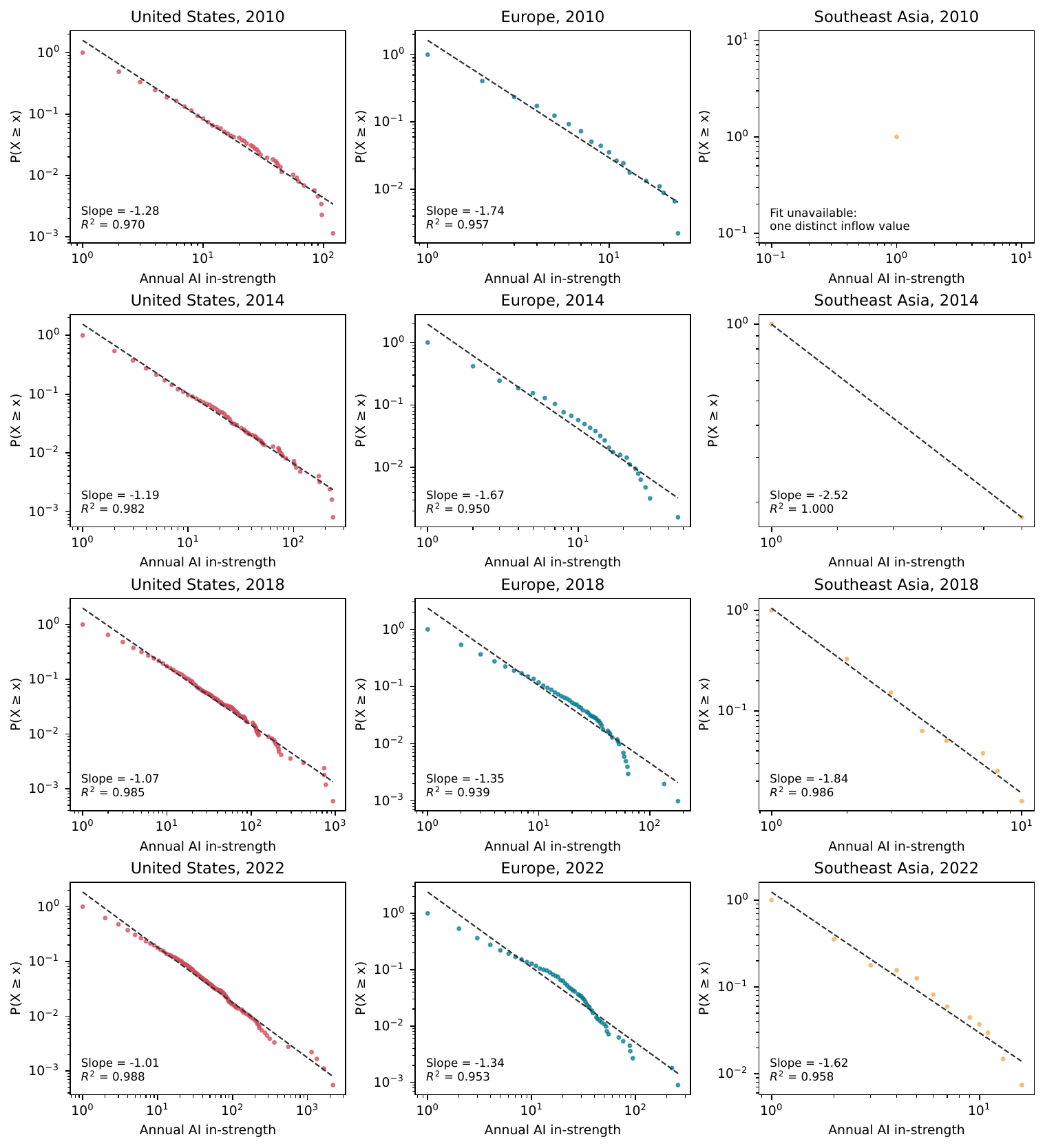}
    \caption{\textbf{Complementary cumulative distribution function of firm-level AI in-strength across markets and years.} Each panel reports the empirical CCDF, $P(X\geq x)$, for firms with positive annual AI in-strength on a log--log scale. Rows correspond to 2010, 2014, 2018 and 2022; columns correspond to the United States, Europe and Southeast Asia. Dashed lines show descriptive ordinary least-squares fits on distinct positive inflow values, with equal weight per value; annotations report slopes and $R^2$. No fit is available for Southeast Asia in 2010 because only one distinct inflow value is observed.}
    \label{fig:ccdf_panel}
\end{figure}

We describe the distribution of annual AI in-strength using complementary cumulative distribution functions (CCDFs) for firms with positive inflows. Let $X$ denote a firm's AI in-strength in a given market-year. At each distinct positive value $x$, the empirical CCDF is
\begin{equation}
    \widehat{\Pr}(X\geq x)=\frac{1}{n}\sum_{i=1}^{n}\mathbbm{1}[X_i\geq x],
\end{equation}
where $n$ is the number of firms with positive AI in-strength. Figure~\ref{fig:ccdf_panel} reports these distributions for the three markets in 2010, 2014, 2018 and 2022. Where at least two distinct inflow values are observed, we fit an ordinary least-squares line to the log-transformed CCDF, weighting each distinct value equally. The slope and $R^2$ summarize the displayed distribution.

The United States and European distributions show extended right tails, with fitted slopes becoming less negative over the selected years. The United States slope changes from $-1.28$ in 2010 to $-1.01$ in 2022, and the European slope from $-1.74$ to $-1.34$. Across these eight panels, $R^2$ ranges from 0.939 to 0.988; the lowest value occurs in Europe in 2018. Southeast Asia has sparse early support. Its 2010 distribution contains eight positive-inflow firms with one distinct inflow value, so no slope is fitted. The 2014 fit uses only two distinct values; its $R^2=1$ therefore provides little evidence about distributional form. In 2018 and 2022, the Southeast Asian slopes are $-1.84$ and $-1.62$, with $R^2$ values of 0.986 and 0.958. 

A less negative fitted slope means that the share of firms exceeding an inflow threshold declines more slowly as that threshold rises. This flattening is consistent with a heavier right tail and relatively greater representation of large inflows. Alongside the rising Gini coefficient and top-decile inflow share, it provides a complementary descriptive indication that AI talent inflows became increasingly unevenly distributed.

\newpage

\section{Descriptive Analysis}

Table~\ref{tab:summary_ai_adoption} presents summary statistics for the main variables used in this study, stratified by whether a firm has ever had an AI-related position in 2010--2022 (AI adopters versus non-adopters). AI-adopting firms are systematically larger and more profitable. They exhibit substantially higher turnover (3,663.9 versus 180.2 million), enterprise value (7,370.0 versus 481.3 million) and total employment (11,749 versus 723). AI adopters also invest more heavily in R\&D (114.6 versus 2.5 million) and hold more cash on average. These differences in financial and firm characteristics are all statistically significant at the $p<0.001$ level.

The two groups also differ in their participation in AI talent flows. AI adopters receive on average 1.65 AI transitions per firm-year, compared with 0.01 for non-adopters, and account for 37,553 of the 38,314 firm-years in the AI network. AI flows are defined by the source position, so a firm without AI-related positions can still receive them when it hires former AI workers into other roles; this occurs in 691 firm-years of 555 non-adopting firms, typically with a single transition. Among firm-years in the AI network, mean PageRank does not differ significantly between the two groups ($p=0.10$), a comparison that rests on only 761 non-adopter firm-years. The sample includes 10,722 AI-adopting firms and 12,874 non-adopting firms.

\begin{table}[!ht]
\centering
\caption{Summary statistics by AI adoption status. Financial values are reported in millions.}
\label{tab:summary_ai_adoption}

\small
\setlength{\tabcolsep}{4pt}
\begin{tabular}{lccccc}
\toprule
& \multicolumn{2}{c}{Non-AI adopters}
& \multicolumn{2}{c}{AI adopters}
& \\
\cmidrule(lr){2-3}
\cmidrule(lr){4-5}
Variable & Mean & SD & Mean & SD & $p$-value \\
\midrule

\addlinespace[0.6em]
\multicolumn{6}{l}{\textbf{Performance}}\\
\hspace{1em}Turnover
& 180.22 & 2,876.77 & 3,663.86 & 16,575.29 & $<0.001$\\
\hspace{1em}Operating revenue
& 180.62 & 3,259.91 & 3,617.08 & 16,430.19 & $<0.001$\\
\hspace{1em}Gross profit
& 75.32 & 2,257.54 & 1,532.93 & 6,394.81 & $<0.001$\\
\hspace{1em}EBITDA
& 32.14 & 1,559.59 & 621.79 & 2,985.80 & $<0.001$\\
\hspace{1em}Net profit
& 4.62 & 219.22 & 240.31 & 1,702.73 & $<0.001$\\
\hspace{1em}Cost of goods sold
& 125.10 & 1,389.47 & 2,337.70 & 12,156.27 & $<0.001$\\
\hspace{1em}Labor expenses
& 19.91 & 430.16 & 395.94 & 1,832.04 & $<0.001$\\
\hspace{1em}Operating expenses
& 58.31 & 1,485.08 & 1,132.03 & 4,845.17 & $<0.001$\\
\hspace{1em}Enterprise value
& 481.31 & 10,812.93 & 7,369.96 & 38,038.95 & $<0.001$\\
\hspace{1em}ROA (\%)
& -4.17 & 20.05 & -0.59 & 17.16 & $<0.001$\\
\hspace{1em}ROE (\%)
& -13.90 & 65.26 & -3.67 & 55.71 & $<0.001$\\

\addlinespace[0.6em]
\multicolumn{6}{l}{\textbf{Firm characteristics}}\\
\hspace{1em}Total assets
& 667.69 & 56,336.85 & 6,615.78 & 44,224.34 & $<0.001$\\
\hspace{1em}Fixed assets
& 220.09 & 18,076.00 & 1,649.22 & 8,418.62 & $<0.001$\\
\hspace{1em}Employment
& 722.66 & 4,302.48 & 11,748.90 & 49,945.70 & $<0.001$\\
\hspace{1em}R\&D expenses
& 2.47 & 19.04 & 114.60 & 903.95 & $<0.001$\\
\hspace{1em}Cash holdings
& 122.64 & 17,550.41 & 694.37 & 10,355.22 & $<0.001$\\
\hspace{1em}Firm age (years)
& 31.81 & 34.10 & 37.29 & 40.87 & $<0.001$\\

\addlinespace[0.6em]
\multicolumn{6}{l}{\textbf{Network centrality}}\\
\hspace{1em}PageRank (AI, $\times10^{3}$)$^{a}$
& 1.35 & 5.63 & 1.01 & 3.54 & 0.102\\
\hspace{1em}In-strength (AI)
& 0.01 & 0.21 & 1.65 & 18.95 & $<0.001$\\
\hspace{1em}Betweenness (AI, $\times10^{3}$)$^{a}$
& 0.04 & 0.30 & 1.02 & 8.33 & $<0.001$\\
\hspace{1em}Net flow (AI)
& 0.01 & 0.17 & 0.00 & 9.74 & 0.645\\
\hspace{1em}PageRank (non-AI, $\times10^{3}$)$^{a}$
& 0.06 & 0.10 & 0.27 & 0.97 & $<0.001$\\
\hspace{1em}In-strength (non-AI)
& 3.58 & 11.11 & 155.88 & 733.96 & $<0.001$\\
\hspace{1em}Betweenness (non-AI, $\times10^{3}$)$^{a}$
& 0.02 & 0.13 & 0.96 & 10.54 & $<0.001$\\
\hspace{1em}Net flow (non-AI)
& 0.20 & 7.68 & -0.13 & 346.88 & 0.740\\

\addlinespace[0.6em]
\multicolumn{6}{l}{\textbf{Sample size}}\\
\hspace{1em}Number of firms
& 12,874 & & 10,722 & & \\
\hspace{1em}Firm-years
& 81,334 & & 125,270 & & \\
\hspace{1em}Firm-years in the AI network
& 761 & & 37,553 & & \\

\bottomrule
\end{tabular}
\par\parbox{\linewidth}{\footnotesize AI adopters are firms with at least one AI-related position in any year from 2010 to 2022; all firm-years of a firm share its label. AI in-strength counts incoming transitions whose source position was AI-related, so a firm without AI-related positions can still receive AI flows when it hires former AI workers into other roles. In-strength and net flow are zero for firm-years outside the corresponding network. $^{a}$PageRank and betweenness are undefined outside the network and are summarized over firm-years in the corresponding network. $p$-values are from Welch two-sample $t$-tests.}
\end{table}

\newpage

\section{Heterogeneity Analysis}

Supplementary Table~\ref{tab:si_heterogeneity} re-estimates the enterprise-value specification within subgroups defined by industry, market, firm size and period, re-pruning each subsample and including subgroup-specific firm and year fixed effects. Point estimates are positive in every subgroup and largest among technology firms (0.047) and in Southeast Asia (0.060), where the sample is small (405 firm-years). Fully interacted models with firm-clustered standard errors do not reject equality of the PageRank coefficient within any partition: technology versus non-technology ($P=0.12$), across the three markets ($P=0.76$), large versus small firms ($P=0.40$) and early versus late periods ($P=0.69$). 

\begin{sidewaystable}
   \caption{\label{tab:si_heterogeneity} Heterogeneity of the AI network-centrality premium: enterprise value}
      \centering
   \small
   \begin{tabular}{lccccccccc}
      \toprule
       & \multicolumn{9}{c}{ln(Enterprise value)}\\
                              & Tech          & Non-Tech     & US           & EU            & SE      & Large          & Small   & Early        & Late \\   
                              & (1)           & (2)          & (3)          & (4)           & (5)     & (6)            & (7)     & (8)          & (9)\\  
      \midrule 
      AI PageRank (log, std.) & 0.047$^{***}$ & 0.018$^{**}$ & 0.026$^{**}$ & 0.028$^{***}$ & 0.060   & 0.020$^{***}$  & 0.006   & 0.016$^{**}$ & 0.007\\   
                              & (0.017)       & (0.008)      & (0.011)      & (0.010)       & (0.047) & (0.007)        & (0.036) & (0.007)      & (0.012)\\   
      Log(1 + AI employees)   & -0.095$^{*}$  & -0.015       & -0.038       & -0.045        & -0.068  & -0.079$^{***}$ & 0.085   & -0.039       & -0.042\\   
                              & (0.050)       & (0.024)      & (0.029)      & (0.034)       & (0.122) & (0.022)        & (0.065) & (0.028)      & (0.036)\\   
      Firms                   &   867         & 2,462        & 1,912        & 1,290         &   127   & 2,611          &   835   & 2,073        & 2,657\\  
       \\
      Observations            & 5,416         & 14,997       & 12,733       & 7,275         & 405     & 17,123         & 2,985   & 9,785        & 9,681\\  
      Within R$^2$            & 0.145         & 0.198        & 0.168        & 0.196         & 0.233   & 0.216          & 0.047   & 0.139        & 0.009\\  
       \\
      Firm fixed effects      & Yes           & Yes          & Yes          & Yes           & Yes     & Yes            & Yes     & Yes          & Yes\\  
      Year fixed effects      & Yes           & Yes          & Yes          & Yes           & Yes     & Yes            & Yes     & Yes          & Yes\\  
      \bottomrule
   \end{tabular}
   
   \parbox{\linewidth}{\footnotesize Dependent variables are measured in year $t+1$ and regressors in year $t$. AI PageRank is log PageRank standardized within market-year. All models control for log(1 + AI employees), log(1 + non-AI employees), log assets and cash/assets. Standard errors clustered by firm are in parentheses. $^{*}P<0.10$, $^{**}P<0.05$, $^{***}P<0.01$. Dependent variable: log enterprise value. Each column re-estimates the model with firm and year fixed effects within the subgroup. Large and small firms are split at the market-year median of total assets; early and late periods are 2010--2017 and 2018--2022. }
\end{sidewaystable}

\FloatBarrier

\section{Robustness Checks}

The association between network position and next-year enterprise value is robust to a broad set of alternative specifications. The fixed-effect, lagged-outcome, centrality, control-set and clustering comparisons hold the sample fixed across their columns, so that differences between columns reflect the specification rather than the observations used; the sample-restriction and end-year checks change the sample by design.

Supplementary Table~\ref{tab:si_robust_fe} replaces calendar-year effects with country-by-year effects, industry-by-year effects or both. The coefficient remains between 0.021 and 0.023 ($P\leq0.005$), so the association is not driven by aggregate shocks to particular countries or industries. Supplementary Table~\ref{tab:si_robust_lagged} adds log enterprise value in year $t-1$ on the sample for which it is observed. The lagged outcome absorbs much of the persistence in valuation (0.294), and the PageRank coefficient falls from 0.017 to 0.012 ($P=0.090$). We report this specification as a sensitivity analysis because lagged dependent variables can introduce bias in short fixed-effects panels.

Supplementary Table~\ref{tab:si_robust_subsample} restricts the sample in five ways. Trimming the smallest and largest 5\% of firms by assets, requiring at least five panel years, retaining only firms that employ AI workers, excluding firms whose standardized raw PageRank ever exceeds the pooled 99th-percentile threshold in the baseline regression sample, and excluding Southeast Asia all yield coefficients between 0.018 and 0.027 ($P\leq0.018$). The positive association persists after excluding firms with exceptionally high standardized PageRank. Supplementary Table~\ref{tab:si_robust_period} varies the last predictor year from 2019 to 2022; the coefficient ranges from 0.016 to 0.023 ($P\leq0.029$).

Supplementary Table~\ref{tab:si_robust_altcent} compares codings of network position on a common sample. The within-market-year percentile rank of PageRank yields a coefficient of 0.050 ($P=0.010$), corresponding to the difference between the bottom and top of the distribution. Percentile ranks of in-strength (0.044, $P=0.042$), closeness (0.041, $P=0.033$) and eigenvector centrality (0.037, $P=0.059$), which, like PageRank, capture embeddedness in the core of the network, are positively associated with valuation. Betweenness, which captures brokerage between otherwise disconnected parts of the network, is negatively associated ($-0.060$, $P=0.009$), suggesting that valuation tracks proximity to the core rather than intermediary positions.

Supplementary Table~\ref{tab:si_robust_controls} shows that the coefficient is essentially unchanged when cash holdings are dropped or when leverage and R\&D intensity are added (0.026--0.027), on the smaller sample for which both are observed. Supplementary Table~\ref{tab:si_robust_cluster} reports standard errors clustered by industry and by firm and year; the coefficient remains significant at the 5\% level under each choice.

\begin{table}[!htbp]
   \caption{\label{tab:si_robust_fe} Robustness: alternative fixed effects}
      \centering
   \small
   \begin{tabular}{lcccc}
      \toprule
       & \multicolumn{4}{c}{ln(Enterprise value)}\\
                                  & Firm + year   & Firm + country-year & Firm + industry-year & Firm + both \\   
                                  & (1)           & (2)                 & (3)                  & (4)\\  
      \midrule 
      AI PageRank (log, std.)     & 0.022$^{***}$ & 0.022$^{***}$       & 0.021$^{***}$        & 0.023$^{***}$\\   
                                  & (0.007)       & (0.007)             & (0.007)              & (0.007)\\   
      Log(1 + AI employees)       & -0.053$^{**}$ & -0.043$^{*}$        & -0.043$^{*}$         & -0.036\\   
                                  & (0.023)       & (0.023)             & (0.023)              & (0.024)\\   
      Firms                       & 3,317         & 3,317               & 3,317                & 3,317\\  
       \\
      Observations                & 20,288        & 20,288              & 20,288               & 20,288\\  
      Within R$^2$                & 0.185         & 0.179               & 0.180                & 0.173\\  
       \\
      Firm fixed effects          & Yes           & Yes                 & Yes                  & Yes\\  
      Year fixed effects          & Yes           & No                  & No                   & No\\  
      Country-year fixed effects  & No            & Yes                 & No                   & Yes\\  
      Industry-year fixed effects & No            & No                  & Yes                  & Yes\\  
      \bottomrule
   \end{tabular}
   
   \parbox{\linewidth}{\footnotesize Dependent variables are measured in year $t+1$ and regressors in year $t$. AI PageRank is log PageRank standardized within market-year. All models control for log(1 + AI employees), log(1 + non-AI employees), log assets and cash/assets. Standard errors clustered by firm are in parentheses. $^{*}P<0.10$, $^{**}P<0.05$, $^{***}P<0.01$.  Dependent variable: log enterprise value. All columns use a common sample with country and industry identifiers. }
\end{table}

\begin{table}[!htbp]
   \caption{\label{tab:si_robust_lagged} Robustness: lagged dependent variable}
      \centering
   \small
   \begin{tabular}{lcc}
      \toprule
       & \multicolumn{2}{c}{ln(Enterprise value)}\\
                              & Baseline       & With lagged DV \\   
                              & (1)            & (2)\\  
      \midrule 
      AI PageRank (log, std.) & 0.017$^{**}$   & 0.012$^{*}$\\   
                              & (0.007)        & (0.007)\\   
      Log(1 + AI employees)   & -0.064$^{***}$ & -0.051$^{***}$\\   
                              & (0.022)        & (0.019)\\   
      Outcome at t-1          &                & 0.293$^{***}$\\   
                              &                & (0.019)\\   
      Firms                   & 2,887          & 2,887\\  
       \\
      Observations            & 17,397         & 17,397\\  
      Within R$^2$            & 0.191          & 0.246\\  
       \\
      Firm fixed effects      & Yes            & Yes\\  
      Year fixed effects      & Yes            & Yes\\  
      \bottomrule
   \end{tabular}
   
   \parbox{\linewidth}{\footnotesize Dependent variables are measured in year $t+1$ and regressors in year $t$. AI PageRank is log PageRank standardized within market-year. All models control for log(1 + AI employees), log(1 + non-AI employees), log assets and cash/assets. Standard errors clustered by firm are in parentheses. $^{*}P<0.10$, $^{**}P<0.05$, $^{***}P<0.01$.  Dependent variable: log enterprise value. Both columns use firm-years with log enterprise value observed in calendar year $t-1$. Firm and year fixed effects are included. }
\end{table}

\begin{table}[!htbp]
   \caption{\label{tab:si_robust_subsample} Robustness: subsample restrictions}
      \centering
   \scriptsize
\setlength{\tabcolsep}{2.5pt}
   \begin{tabular}{lccccc}
      \toprule
       & \multicolumn{5}{c}{ln(Enterprise value)}\\
                              & Trim assets 5\%  & 5+ panel years & AI-active firms & Drop top-1\% PageRank firms  & Excl. SE \\   
                              & (1)              & (2)            & (3)             & (4)                          & (5)\\  
      \midrule 
      AI PageRank (log, std.) & 0.027$^{***}$    & 0.018$^{**}$   & 0.023$^{***}$   & 0.024$^{***}$                & 0.022$^{***}$\\   
                              & (0.008)          & (0.008)        & (0.007)         & (0.008)                      & (0.008)\\   
      Log(1 + AI employees)   & -0.062$^{***}$   & -0.101$^{***}$ & -0.054$^{**}$   & -0.052$^{**}$                & -0.050$^{**}$\\   
                              & (0.023)          & (0.026)        & (0.022)         & (0.023)                      & (0.023)\\   
      Firms                   & 2,972            & 1,782          & 3,309           & 3,284                        & 3,202\\  
       \\
      Observations            & 18,290           & 16,180         & 20,367          & 19,897                       & 20,008\\  
      Within R$^2$            & 0.199            & 0.240          & 0.186           & 0.180                        & 0.185\\  
       \\
      Firm fixed effects      & Yes              & Yes            & Yes             & Yes                          & Yes\\  
      Year fixed effects      & Yes              & Yes            & Yes             & Yes                          & Yes\\  
      \bottomrule
   \end{tabular}
   
   \parbox{\linewidth}{\footnotesize Dependent variables are measured in year $t+1$ and regressors in year $t$. AI PageRank is log PageRank standardized within market-year. All models control for log(1 + AI employees), log(1 + non-AI employees), log assets and cash/assets. Standard errors clustered by firm are in parentheses. $^{*}P<0.10$, $^{**}P<0.05$, $^{***}P<0.01$.  Dependent variable: log enterprise value. Trimming removes the bottom and top 5\% of total assets; AI-active firms employ AI workers in at least one year; For the top-1\% exclusion, raw PageRank is standardized within market-year among Orbis-matched firms; the 99th-percentile threshold is then calculated over the pooled baseline regression sample, and firms exceeding it in any baseline-sample year are dropped.} Firm and year fixed effects are included. 
\end{table}

\begin{table}[!htbp]
   \caption{\label{tab:si_robust_period} Robustness: sample-period sensitivity}
   \bigskip
   \centering
   \small
   \begin{tabular}{lcccc}
      \toprule
       & \multicolumn{4}{c}{ln(Enterprise value)}\\
                              & Through 2019 & Through 2020  & Through 2021  & Through 2022 \\   
                              & (1)          & (2)           & (3)           & (4)\\  
      \midrule 
      AI PageRank (log, std.) & 0.016$^{**}$ & 0.020$^{***}$ & 0.019$^{**}$  & 0.023$^{***}$\\   
                              & (0.007)      & (0.007)       & (0.008)       & (0.007)\\   
      Log(1 + AI employees)   & -0.039       & -0.051$^{**}$ & -0.053$^{**}$ & -0.052$^{**}$\\   
                              & (0.026)      & (0.024)       & (0.023)       & (0.022)\\   
      Firms                   & 2,566        & 2,805         & 3,144         & 3,329\\  
       \\
      Observations            & 13,787       & 15,887        & 18,349        & 20,413\\  
      Within R$^2$            & 0.203        & 0.220         & 0.188         & 0.185\\  
       \\
      Firm fixed effects      & Yes          & Yes           & Yes           & Yes\\  
      Year fixed effects      & Yes          & Yes           & Yes           & Yes\\  
      \bottomrule
   \end{tabular}
   
   \parbox{\linewidth}{\footnotesize Dependent variables are measured in year $t+1$ and regressors in year $t$. AI PageRank is log PageRank standardized within market-year. All models control for log(1 + AI employees), log(1 + non-AI employees), log assets and cash/assets. Standard errors clustered by firm are in parentheses. $^{*}P<0.10$, $^{**}P<0.05$, $^{***}P<0.01$.  Dependent variable: log enterprise value. Columns restrict predictor years to 2010 through the stated end year. Firm and year fixed effects are included. }
\end{table}

\begin{table}[!htbp]
   \caption{\label{tab:si_robust_altcent} Robustness: PageRank coding and alternative centralities}
   \bigskip
   \centering
   \small
   \begin{tabular}{lcccccc}
      \toprule
       & \multicolumn{6}{c}{ln(Enterprise value)}\\
                               & (1)           & (2)           & (3)           & (4)           & (5)           & (6)\\  
      \midrule 
      AI PageRank (log, std.)  & 0.023$^{***}$ &               &               &               &               &   \\   
                               & (0.007)       &               &               &               &               &   \\   
      AI PageRank (percentile) &               & 0.050$^{***}$ &               &               &               &   \\   
                               &               & (0.019)       &               &               &               &   \\   
      In-strength (percentile) &               &               & 0.044$^{**}$  &               &               &   \\   
                               &               &               & (0.022)       &               &               &   \\   
      Closeness (percentile)   &               &               &               & 0.041$^{**}$  &               &   \\   
                               &               &               &               & (0.019)       &               &   \\   
      Eigenvector (percentile) &               &               &               &               & 0.037$^{*}$   &   \\   
                               &               &               &               &               & (0.020)       &   \\   
      Betweenness (percentile) &               &               &               &               &               & -0.060$^{***}$\\   
                               &               &               &               &               &               & (0.023)\\   
      Log(1 + AI employees)    & -0.052$^{**}$ & -0.049$^{**}$ & -0.049$^{**}$ & -0.049$^{**}$ & -0.048$^{**}$ & -0.041$^{*}$\\   
                               & (0.022)       & (0.022)       & (0.022)       & (0.022)       & (0.022)       & (0.023)\\   
      Firms                    & 3,329         & 3,329         & 3,329         & 3,329         & 3,329         & 3,329\\  
       \\
      Observations             & 20,413        & 20,413        & 20,413        & 20,413        & 20,413        & 20,413\\  
      Within R$^2$             & 0.185         & 0.185         & 0.185         & 0.185         & 0.185         & 0.185\\  
       \\
      Firm fixed effects       & Yes           & Yes           & Yes           & Yes           & Yes           & Yes\\  
      Year fixed effects       & Yes           & Yes           & Yes           & Yes           & Yes           & Yes\\  
      \bottomrule
   \end{tabular}
   
   \parbox{\linewidth}{\footnotesize Dependent variables are measured in year $t+1$ and regressors in year $t$. AI PageRank is log PageRank standardized within market-year. All models control for log(1 + AI employees), log(1 + non-AI employees), log assets and cash/assets. Standard errors clustered by firm are in parentheses. $^{*}P<0.10$, $^{**}P<0.05$, $^{***}P<0.01$.  Dependent variable: log enterprise value. All columns use a common sample. Percentile measures rank firms within market-year. Firm and year fixed effects are included. }
\end{table}

\begin{table}[!htbp]
   \caption{\label{tab:si_robust_controls} Robustness: control sets}
   \bigskip
   \centering
   \small
   \begin{tabular}{lccc}
      \toprule
       & \multicolumn{3}{c}{ln(Enterprise value)}\\
                              & Inputs only    & Baseline       & Add leverage and R\&D \\    
                              & (1)            & (2)            & (3)\\  
      \midrule 
      AI PageRank (log, std.) & 0.026$^{***}$  & 0.027$^{***}$  & 0.027$^{***}$\\   
                              & (0.009)        & (0.009)        & (0.009)\\   
      Log(1 + AI employees)   & -0.100$^{***}$ & -0.099$^{***}$ & -0.100$^{***}$\\   
                              & (0.028)        & (0.028)        & (0.028)\\   
      Firms                   & 2,351          & 2,351          & 2,351\\  
       \\
      Observations            & 13,884         & 13,884         & 13,884\\  
      Within R$^2$            & 0.196          & 0.197          & 0.197\\  
       \\
      Firm fixed effects      & Yes            & Yes            & Yes\\  
      Year fixed effects      & Yes            & Yes            & Yes\\  
      \bottomrule
   \end{tabular}
   
   \parbox{\linewidth}{\footnotesize Dependent variable: log enterprise value. AI PageRank is log PageRank standardized within market-year. All columns use the sample with leverage and R\&D intensity observed. Inputs only: log(1 + AI employees), log(1 + non-AI employees) and log assets; baseline adds cash/assets. Firm and year fixed effects; standard errors clustered by firm. $^{*}P<0.10$, $^{**}P<0.05$, $^{***}P<0.01$. }
\end{table}

\begin{table}[!htbp]
   \caption{\label{tab:si_robust_cluster} Robustness: clustering}
      \centering
   \small
   \begin{tabular}{lccc}
      \toprule
       & \multicolumn{3}{c}{ln(Enterprise value)}\\
                              & Firm          & Industry                & Firm and year \\   
                              & (1)           & (2)                     & (3)\\  
      \midrule 
      AI PageRank (log, std.) & 0.023$^{***}$ & 0.023$^{***}$           & 0.023$^{**}$\\   
                              & (0.007)       & (0.006)                 & (0.008)\\   
      Log(1 + AI employees)   & -0.052$^{**}$ & -0.052$^{**}$           & -0.052$^{*}$\\   
                              & (0.022)       & (0.023)                 & (0.025)\\   
      Firms                   & 3,329         & 3,329                   & 3,329\\  
       \\
      Standard-Errors         & Firm          & Industry (NACE section) & Firm \& Year \\    
      Observations            & 20,413        & 20,413                  & 20,413\\  
      Within R$^2$            & 0.185         & 0.185                   & 0.185\\  
       \\
      Firm fixed effects      & Yes           & Yes                     & Yes\\  
      Year fixed effects      & Yes           & Yes                     & Yes\\  
      \bottomrule
   \end{tabular}
   
   \parbox{\linewidth}{\footnotesize Dependent variable: log enterprise value. AI PageRank is log PageRank standardized within market-year. Coefficients are identical across columns; only the clustering of standard errors changes (firm, NACE main section, or firm and year). Firm and year fixed effects and baseline controls are included. $^{*}P<0.10$, $^{**}P<0.05$, $^{***}P<0.01$. }
\end{table}

\FloatBarrier

\section{Event-Study Design Details}
\label{si:eventstudy}

\paragraph{Event definition.} Within each market-year, we rank finite AI PageRank values using average ranks for ties, divided by the number of firms. The jump is the current rank minus its maximum over the preceding three exact calendar years. Its threshold is the 90th percentile among history-eligible firm-years in 2013--2019, calculated before financial filtering (0.1127 in rank units). A qualifying event also requires the preceding maximum minus the year-$-1$ rank to be at most 0.10. We detect each firm's first event in the available 2010--2022 panel and retain first events in 2013--2019. No post-event persistence or employment-stability condition is imposed. There are 456 detected events in this interval; 246 meet the pre-event eligibility requirements and 139 are matched.

\paragraph{Matching.} Treated and control candidates must have complete pre-event ranks at $-3,-2,-1$; AI employment, assets and enterprise value at $-1$; AI-employment and enterprise-value changes from $-3$ to $-1$; and year-$-1$ non-AI employment and industry. Controls have no qualifying event detected through 2022. Matching is exact on market, event year and year-$-1$ NACE main section, with an absolute pairwise caliper of 0.20 on each pre-event rank. The distance uses eight variables: the three ranks, log one plus AI employment and its two-year change, log assets, log enterprise value and its two-year change. We standardize using the complete eligible 2013--2019 reference sample and apply a Mahalanobis distance, adding $10^{-6}$ to the standardized covariance diagonal. The square root of squared distance divided by eight must not exceed 1.5. We select the nearest control, breaking ties by firm identifier; controls may be reused across events. A parallel specification, reported alongside the main estimates, selects up to three nearest controls, each weighted by one over the number of controls, and holds treated event-time support fixed to the one-to-one sample. Of the 246 eligible events, 139 found a match (110 in the United States and 29 in Europe), served by 105 distinct control firms; no Southeast Asian event had an eligible control within the industry and rank constraints. Supplementary Table~\ref{tab:si_event_balance} reports standardized mean differences in the enterprise-value sample, which are below 0.17 in absolute value for all matching variables under one-to-one matching.

\paragraph{Estimation.} For each outcome, both firms must have all three pre-event observations and contribute a baseline and at least one post-event comparison. We retain finite outcomes and event-time cells containing treated and control observations, without requiring complete post-event histories. Treatment and total control weight each equal one per event-time cell. Financial amount logs require positive values. We estimate
\begin{equation}
y_{i,e,\tau}=\sum_{k\ne-1}\beta_k\,Treat_{i,e}\mathbbm{1}[\tau=k]+\sum_{k\ne-1}\mathbbm{1}[\tau=k]\,X_{i,e}'\gamma_k+\mu_{i\times e}+\lambda_{e\times\tau}+\varepsilon_{i,e,\tau},
\end{equation}
for $\tau\in[-3,5]$, where $X_{i,e}$ contains year-$-1$ log one plus AI employment, log one plus non-AI employment and log assets, with firm-by-event and event-by-relative-year fixed effects, year $-1$ omitted, and firm-clustered inference. Because the panel ends in 2022, year 4 is observed only for events in 2013--2018 and year 5 only for events in 2013--2017. A separate regression replaces the treatment interactions with $\theta Treat_{i,e}\mathbbm{1}[\tau\geq0]$ to estimate the pooled post-event association. This coefficient is not an arithmetic average of dynamic estimates. Joint Wald tests assess the two pre-event coefficients. Sensitivity estimates use three or five never-treated controls, holding treated event-time support fixed to the one-to-one sample separately for each outcome, or one or three not-yet-treated controls, which leave the comparison group from their own event year; each is estimated without baseline adjustment, with the main adjustment, and with the cash ratio added. We also estimate the jump in log PageRank standardized within market-year to compare its size with the panel regressor. All 16 estimated outcomes are reported below.

\paragraph{Enterprise value.} In the main specification, the pre-event coefficients for enterprise value are close to zero (joint $P=0.75$). The event-year coefficient is 0.015 (95\% CI $-0.061$ to 0.092), and the coefficients for years 1 to 5 are 0.103, 0.142, 0.112, 0.181 and 0.233 log points; those for years 2, 4 and 5 have $P<0.05$, and those for years 1 and 3 have $P=0.068$ and $P=0.116$. The pooled post-event coefficient is 0.100 (s.e.\ 0.048, $P=0.036$), and the average of the six post-event coefficients is 0.131 ($P=0.011$). Main-text Fig.~4 also reports the specification with three never-treated controls per event on the same treated support. Its enterprise-value estimates are 0.026 in the event year and 0.112, 0.171, 0.142, 0.204 and 0.230 in years 1 to 5, each with $P<0.05$; the pooled coefficient is 0.121 (s.e.\ 0.042, $P=0.004$) and the six-year average 0.148 ($P=0.001$). The pre-event coefficients are $-0.004$ and 0.043 (joint $P=0.074$), and the largest absolute standardized difference in the matching variables is 0.23, compared with 0.17 under one-to-one matching, mainly because the additional controls have lower pre-event enterprise value and assets.

\paragraph{Income statement and costs.} Supplementary Table~\ref{tab:si_event_all} reports every outcome estimated under the main specification. Top-line measures rose with enterprise value: the pooled differences are 0.084 log points for operating revenue ($P=0.039$) and 0.089 for turnover ($P=0.034$). Gross profit (0.072, $P=0.093$) and EBITDA (0.093, $P=0.20$) are positive but imprecise, and operating and net profit are imprecise and observed for fewer events (96 and 81), with weaker pre-event balance (joint pre-event $P=0.04$ and 0.09). Cost levels are imprecise: cost of goods sold rises by 0.065 log points ($P=0.15$) and other operating expenses by 0.004 ($P=0.91$). Relative to turnover, cost of goods sold ($-0.001$), staff costs (0.010) and other operating expenses ($-0.017$) show no detectable difference in any post-event window. Staff costs are reported for only 44 events. R\&D expenditure, reported for 80 events, rises by 0.143 log points ($P=0.008$) and is positive in both the early and late post-event windows. With three never-treated controls per event, the pooled R\&D difference is 0.136 log points ($P=0.004$; joint pre-event $P=0.43$).

\paragraph{Sensitivity.} We examined robustness using alternative comparison groups and baseline adjustments (Supplementary Table~\ref{tab:si_event_sensitivity}). These specifications varied the number of never-treated controls, used not-yet-treated firms as an alternative comparison group, and included different sets of baseline characteristics interacted with relative year. Never-treated specifications held treated event-time support fixed within each outcome; not-yet-treated controls left the comparison group when their own event occurred. Enterprise-value and operating-revenue estimates remained positive across these alternative specifications, consistent with the main findings. Evidence for gross-profit and EBITDA gains remained less conclusive.

\paragraph{Comparison with the panel estimates.} To compare magnitudes, we re-estimate the main specification with the panel regressor, log AI PageRank standardized within market-year among Orbis-matched firms, as the outcome (137 events). The jump raises this measure by 0.87 standard deviations in the event year (s.e.\ 0.07) and by 0.29 to 0.44 standard deviations in years 1 to 5; the pooled post-event difference is 0.45 (s.e.\ 0.05). At the panel coefficient of 0.023 log points per standard deviation, these changes correspond to enterprise-value differences of about 0.02 log points in the year after the event and about 0.01 thereafter, compared with event-study estimates of 0.10 to 0.23 in years 1 to 5. The event firms' AI employment and operating revenue also rise after the jump (main-text Fig.~4b,d). The event study thus captures valuation changes accompanying a broader expansion in network position, AI employment and business activity.

\begin{table}[!htbp]
\caption{\label{tab:si_event_balance}Covariate balance of matched event and control firms}
\centering
\small
\begin{tabular}{lccccc}
\toprule
 & \multicolumn{3}{c}{Never-treated controls} & \multicolumn{2}{c}{Not-yet-treated controls}\\
\cmidrule(lr){2-4}\cmidrule(lr){5-6}
 & 1:1 & 1:3 & 1:5 & 1:1 & 1:3\\
\midrule
PageRank percentile, $t-1$ & -0.109 & -0.083 & -0.075 & -0.049 & -0.042\\
PageRank percentile, $t-2$ & -0.106 & -0.115 & -0.112 & -0.089 & -0.087\\
PageRank percentile, $t-3$ & -0.170 & -0.145 & -0.138 & -0.145 & -0.128\\
Log(1 + AI employees), $t-1$ & 0.018 & 0.105 & 0.141 & 0.053 & 0.133\\
Change in log(1 + AI employees), $t-3$ to $t-1$ & -0.116 & -0.027 & -0.044 & -0.091 & -0.034\\
Log assets, $t-1$ & 0.069 & 0.151 & 0.187 & 0.102 & 0.191\\
Log enterprise value, $t-1$ & 0.151 & 0.226 & 0.260 & 0.153 & 0.225\\
Change in log enterprise value, $t-3$ to $t-1$ & -0.032 & -0.014 & -0.037 & -0.011 & -0.036\\
\bottomrule
\end{tabular}
\par\parbox{\linewidth}{\footnotesize Standardized mean differences (treated minus weighted controls) at the baseline year, scaled by the standard deviation in the eligible 2013--2019 reference sample, for the enterprise-value estimation sample. The eight variables enter the Mahalanobis matching distance.}
\end{table}

\begin{table}[!htbp]
\caption{\label{tab:si_event_sensitivity}Matched event study: pooled post-event estimates by comparison group and pre-event adjustment}
\centering
\small
\begin{tabular}{lccccc}
\toprule
 & \multicolumn{3}{c}{Never-treated controls} & \multicolumn{2}{c}{Not-yet-treated controls}\\
\cmidrule(lr){2-4}\cmidrule(lr){5-6}
 & 1:1 & 1:3 & 1:5 & 1:1 & 1:3\\
\midrule
\multicolumn{6}{l}{\textit{A. No baseline adjustment}}\\
AI PageRank percentile & 0.141$^{***}$ & 0.150$^{***}$ & 0.149$^{***}$ & 0.172$^{***}$ & 0.171$^{***}$\\
 & (0.015) & (0.013) & (0.012) & (0.016) & (0.012)\\
AI employment & 0.083$^{**}$ & 0.105$^{***}$ & 0.103$^{***}$ & 0.073$^{*}$ & 0.089$^{***}$\\
 & (0.041) & (0.034) & (0.033) & (0.038) & (0.029)\\
Enterprise value & 0.089$^{*}$ & 0.104$^{**}$ & 0.104$^{**}$ & 0.070 & 0.067$^{*}$\\
 & (0.048) & (0.044) & (0.043) & (0.044) & (0.040)\\
Operating revenue (OPRE) & 0.077$^{*}$ & 0.074$^{**}$ & 0.065$^{*}$ & 0.071$^{*}$ & 0.051$^{*}$\\
 & (0.041) & (0.034) & (0.033) & (0.037) & (0.031)\\
Gross profit (GROS) & 0.068 & 0.061 & 0.056 & 0.056 & 0.030\\
 & (0.043) & (0.037) & (0.037) & (0.038) & (0.032)\\
EBITDA & 0.094 & 0.091 & 0.082 & 0.071 & 0.074\\
 & (0.072) & (0.075) & (0.072) & (0.055) & (0.060)\\
\midrule
\multicolumn{6}{l}{\textit{B. Baseline scale controls $\times$ relative year (main specification)}}\\
AI PageRank percentile & 0.139$^{***}$ & 0.138$^{***}$ & 0.136$^{***}$ & 0.166$^{***}$ & 0.158$^{***}$\\
 & (0.014) & (0.013) & (0.012) & (0.015) & (0.012)\\
AI employment & 0.077$^{*}$ & 0.101$^{***}$ & 0.097$^{***}$ & 0.071$^{*}$ & 0.085$^{***}$\\
 & (0.040) & (0.033) & (0.033) & (0.038) & (0.029)\\
Enterprise value & 0.100$^{**}$ & 0.121$^{***}$ & 0.120$^{***}$ & 0.073$^{*}$ & 0.084$^{**}$\\
 & (0.048) & (0.042) & (0.041) & (0.041) & (0.039)\\
Operating revenue (OPRE) & 0.084$^{**}$ & 0.092$^{***}$ & 0.085$^{***}$ & 0.077$^{**}$ & 0.069$^{**}$\\
 & (0.040) & (0.032) & (0.032) & (0.035) & (0.029)\\
Gross profit (GROS) & 0.072$^{*}$ & 0.077$^{**}$ & 0.075$^{**}$ & 0.057 & 0.048\\
 & (0.042) & (0.036) & (0.036) & (0.037) & (0.031)\\
EBITDA & 0.093 & 0.090 & 0.077 & 0.060 & 0.070\\
 & (0.071) & (0.067) & (0.065) & (0.054) & (0.055)\\
\midrule
\multicolumn{6}{l}{\textit{C. Baseline scale controls and cash ratio $\times$ relative year}}\\
AI PageRank percentile & 0.139$^{***}$ & 0.135$^{***}$ & 0.132$^{***}$ & 0.169$^{***}$ & 0.157$^{***}$\\
 & (0.014) & (0.013) & (0.012) & (0.015) & (0.012)\\
AI employment & 0.074$^{*}$ & 0.092$^{***}$ & 0.090$^{***}$ & 0.069$^{*}$ & 0.076$^{***}$\\
 & (0.040) & (0.034) & (0.034) & (0.037) & (0.029)\\
Enterprise value & 0.096$^{**}$ & 0.114$^{***}$ & 0.112$^{***}$ & 0.065 & 0.076$^{*}$\\
 & (0.048) & (0.043) & (0.042) & (0.040) & (0.039)\\
Operating revenue (OPRE) & 0.087$^{**}$ & 0.088$^{***}$ & 0.080$^{**}$ & 0.074$^{**}$ & 0.066$^{**}$\\
 & (0.040) & (0.033) & (0.032) & (0.034) & (0.029)\\
Gross profit (GROS) & 0.075$^{*}$ & 0.070$^{*}$ & 0.066$^{*}$ & 0.052 & 0.042\\
 & (0.041) & (0.037) & (0.036) & (0.036) & (0.031)\\
EBITDA & 0.108 & 0.096 & 0.083 & 0.070 & 0.077\\
 & (0.069) & (0.066) & (0.064) & (0.052) & (0.054)\\
\midrule
Enterprise-value events & 136 & 136 & 136 & 156 & 156\\
\bottomrule
\end{tabular}
\par\parbox{\linewidth}{\footnotesize Entries are pooled post-event coefficients $\theta$ (relative years 0 to 5 versus $-3$ to $-1$) from stacked regressions with firm-by-event and event-by-relative-year fixed effects; standard errors clustered by firm in parentheses. Never-treated columns hold treated event-time support fixed at the one-to-one sample for each outcome; not-yet-treated columns use their own matched samples, with controls censored from their own event year. Scale controls are year-$-1$ log(1 + AI employees), log(1 + non-AI employees) and log assets. Financial outcomes are positive-value logs. $^{*}P<0.10$, $^{**}P<0.05$, $^{***}P<0.01$.}
\end{table}

\begin{table}[!htbp]
\caption{\label{tab:si_event_all}Matched event study: all estimated outcomes under the main specification}
\centering
\footnotesize
\setlength{\tabcolsep}{4pt}
\begin{tabular}{lccccc}
\toprule
 & Pooled $\theta$ & Years 0--2 & Years 3--5 & Years 0--5 & Pre-event $P$\\
\midrule
AI PageRank percentile & 0.139$^{***}$ & 0.135$^{***}$ & 0.120$^{***}$ & 0.127$^{***}$ & 0.24\\
 & (0.014) & (0.017) & (0.021) & (0.015) & \\
AI employment & 0.077$^{*}$ & 0.052$^{*}$ & 0.129$^{**}$ & 0.090$^{**}$ & 0.39\\
 & (0.040) & (0.028) & (0.052) & (0.037) & \\
Enterprise value & 0.100$^{**}$ & 0.087$^{*}$ & 0.175$^{**}$ & 0.131$^{**}$ & 0.75\\
 & (0.048) & (0.046) & (0.073) & (0.051) & \\
Operating revenue (OPRE) & 0.084$^{**}$ & 0.079$^{***}$ & 0.087$^{*}$ & 0.083$^{**}$ & 0.91\\
 & (0.040) & (0.030) & (0.052) & (0.038) & \\
Turnover (TURN) & 0.089$^{**}$ & 0.091$^{**}$ & 0.101$^{*}$ & 0.096$^{**}$ & 0.91\\
 & (0.042) & (0.038) & (0.058) & (0.045) & \\
Gross profit (GROS) & 0.072$^{*}$ & 0.072$^{**}$ & 0.057 & 0.064 & 0.88\\
 & (0.042) & (0.030) & (0.056) & (0.039) & \\
EBITDA & 0.093 & 0.111$^{*}$ & 0.051 & 0.081 & 0.44\\
 & (0.071) & (0.058) & (0.097) & (0.068) & \\
Operating profit & 0.031 & 0.035 & 0.101 & 0.068 & 0.04\\
 & (0.090) & (0.085) & (0.129) & (0.100) & \\
Net profit & 0.098 & 0.045 & 0.149 & 0.097 & 0.09\\
 & (0.118) & (0.118) & (0.150) & (0.125) & \\
Cost of goods sold & 0.065 & 0.057$^{**}$ & 0.072 & 0.065$^{*}$ & 0.94\\
 & (0.045) & (0.024) & (0.059) & (0.039) & \\
Staff costs & 0.081 & 0.020 & 0.048 & 0.034 & 0.59\\
 & (0.088) & (0.090) & (0.130) & (0.104) & \\
Other operating expenses & 0.004 & 0.022 & 0.004 & 0.013 & 0.50\\
 & (0.039) & (0.029) & (0.060) & (0.039) & \\
R\&D expenditure & 0.143$^{***}$ & 0.110$^{**}$ & 0.151$^{**}$ & 0.130$^{**}$ & 0.26\\
 & (0.053) & (0.044) & (0.074) & (0.053) & \\
Cost of goods sold / turnover & -0.001 & -0.005 & 0.002 & -0.002 & 0.80\\
 & (0.007) & (0.005) & (0.008) & (0.006) & \\
Staff costs / turnover & 0.010 & -0.010 & 0.018 & 0.004 & 0.18\\
 & (0.009) & (0.009) & (0.020) & (0.013) & \\
Other operating expenses / turnover & -0.017 & -0.021 & 0.009 & -0.006 & 0.71\\
 & (0.016) & (0.016) & (0.024) & (0.018) & \\
\bottomrule
\end{tabular}
\par\parbox{\linewidth}{\footnotesize One-to-one never-treated matching with year-$-1$ log(1 + AI employees), log(1 + non-AI employees) and log assets interacted with relative year; firm-by-event and event-by-relative-year fixed effects; standard errors clustered by firm in parentheses. Pooled $\theta$ is from a separate regression on $Treat\times\mathbbm{1}[\tau\geq0]$; year averages are equal-weight averages of the dynamic coefficients. Pre-event $P$ is the joint test of years $-3$ and $-2$. Amounts are positive-value logs; ratios are relative to turnover. $^{*}P<0.10$, $^{**}P<0.05$, $^{***}P<0.01$.}
\end{table}

\FloatBarrier

\bibliography{sn-bibliography}